\documentstyle[aaspp4,tighten,flushrt]{article}
\newcommand{\beq}{\begin{equation}}
\newcommand{\eeq}{\end{equation}}
\newcommand{\beqa}{\begin{eqnarray}}
\newcommand{\eeqa}{\end{eqnarray}}
\newcommand{\bml}{\begin{mathletters}}
\newcommand{\eml}{\end{mathletters}}

\newcommand{\lexp}{\mathop{\langle}}
\newcommand{\rexp}{\mathop{\rangle}}
\newcommand{\rexpc}{\mathop{\rangle_c}}

\def\d{\delta}
\def\te{\theta}

\def\dt{\tilde \delta}
\def\dD{[\delta_{\rm D}]}
\def\del{{\bf \nabla}}

\font\BF=cmmib10
\font\BFs=cmmib10 scaled 833
\def\k{{\hbox{\BF k}}}
\def\x{{\hbox{\BF x}}}

\def\ks{{\hbox{\BFs k}}}
\def\xs{{\hbox{\BFs x}}}
\def\q{{\hbox{\BF q}}}
\def\v{{\hbox{\BF v}}}

\def\la{\mathrel{\mathpalette\fun <}}
\def\ga{\mathrel{\mathpalette\fun >}}
\def\fun#1#2{\lower3.6pt\vbox{\baselineskip0pt\lineskip.9pt
        \ialign{$\mathsurround=0pt#1\hfill##\hfil$\crcr#2\crcr\sim\crcr}}}

\begin{document}
%%%%%%%%%%%%%%%%%%%%%%%%%%%%%%%%%%%%%%%%%%%%%%%%%%%%%%%%%%%%%%%%%%%%%%%%%%%%%%%%%%%%%%

\hfill{\small CITA-97-42}
\normalsize

%
%
%==========================================================================
\title{Transients from Initial Conditions: A Perturbative Analysis}
%==========================================================================
%
%
\author{Rom\'{a}n Scoccimarro\footnote{\sf scoccima@cita.utoronto.ca}}

\vskip 1pc

\affil{CITA, McLennan Physical Labs, 60 St George Street,
Toronto ON M5S 3H8, Canada}

%
%================
\begin{abstract}
%================
%

The standard procedure to generate initial conditions in numerical
simulations of structure formation is to use the  Zel'dovich
approximation (ZA). Although the ZA
correctly reproduces the linear growing modes of density and velocity
perturbations, non-linear growth is inaccurately
represented, particularly for velocity perturbations because of the
ZA failure to conserve momentum.  
This implies that it takes time for the actual dynamics
to establish the correct statistical properties of density and 
velocity fields. 

We extend the standard formulation of non-linear 
perturbation theory (PT) to include transients as 
non-linear excitations of decaying modes caused by the initial
conditions. These new non-linear solutions interpolate between the
initial conditions and the late-time solutions 
given by the exact non-linear dynamics.
To quantify the magnitude of transients, we focus on higher-order
statistics of the density contrast $\d$ and velocity divergence
$\Theta$, characterized by the $S_p$ and $T_p$ parameters. These
describe the non-Gaussianity of 
the probability distribution through its connected moments 
$\lexp \d^p \rexpc \equiv S_p \lexp \d^2 \rexp^{p-1}$, 
$\lexp \Theta^p\rexpc \equiv T_p \lexp \Theta^2 \rexp^{p-1}$. 
We calculate $S_p(a)$ and $T_p(a)$  to leading-order in PT with 
top-hat smoothing as a function of scale factor $a$.  

We find that the time-scale of transients 
is determined, at a given order $p$, by the effective spectral index
$n_{\rm eff}$. The skewness
factor $S_3$ ($T_3$) attains 10$\%$ accuracy only after $a \approx 6$ ($a
\approx 15$) for $n_{\rm eff} \approx 0$, whereas higher (lower)
$n_{\rm eff}$ demands more (less) expansion away from the initial
conditions.  These requirements   
become much more stringent as $p$ increases, always showing slower
decay of transients for $T_p$ than $S_p$. 
For models with density parameter $\Omega \neq 
1$, the conditions above apply to the linear growth factor, thus an $\Omega=0.3$
open model requires roughly a factor of two larger expansion than a
critical density model to reduce transients by the 
same amount. The predicted transients in $S_p$ are in good   
agreement with numerical simulations. 

More accurate initial conditions can be achieved  by using 
second-order Lagrangian PT (2LPT), which reproduces growing
modes up to second-order and thus eliminates transients in the
skewness parameters. We show that for $p>3$ this scheme can reduce the
required expansion by more than an order of magnitude compared to the
ZA. Setting up 2LPT initial conditions only requires minimal, 
inexpensive changes to ZA codes. We suggest  simple steps
for its implementation.

\end{abstract}

\clearpage 

%
%
%=======================
\section{Introduction}
%=======================
%
%

Gravitational instability is widely considered responsible for the
formation and evolution of large-scale structures in the Universe. 
The non-linear nature of gravitational
clustering makes analytical calculations only possible for models with
restrictive symmetries or in cases where the density fluctuations are
small enough that a perturbative approach is possible. For this
reason, numerical simulations are a vital resource for understanding how
large-scale structures form and evolve in the Universe.

The standard procedure in numerical simulations is to set up the
initial perturbations, assumed to be Gaussian, by using the
Zel'dovich (1970) approximation 
(Klypin \& Shandarin 1983; see also Efstathiou et al. 1985, hereafter
EDFW). This 
gives a useful prescription to perturb the positions of
particles from some regular pattern (commonly a grid or a ``glass'')
and assign  them velocities according to the growing mode in linear
perturbation theory. In this way, one can generate fluctuations 
with any desired power spectrum and then numerically evolve them
forward in time to the present epoch.

Although the
Zel'dovich approximation (hereafter ZA) correctly reproduces the
linear growing modes of density and velocity perturbations, non-linear
correlations are known to be inaccurate when compared to the exact
dynamics (Grinstein \& Wise 1987, Juszkiewicz, Bouchet \& Colombi
1993, Bernardeau 1994, Catelan \& Moscardini 1994, Juszkiewicz et
al. 1995).  This implies that it may take a 
non-negligible amount of time for the exact dynamics to establish
the correct statistical properties of density and velocity
fields. This transient behavior affects in greater extent statistical
quantities which are sensitive to phase correlations of density and
velocity fields; by contrast, the two-point function, variance, and 
power spectrum of density fluctuations at large scales can be
described by linear perturbation theory, and are thus unaffected by the
incorrect higher-order correlations imposed by the initial conditions.

In this work, we therefore concentrate on higher-order statistics  
such as the one-point cumulants of density and velocity divergence
fields, characterized by the so-called $S_p$ and $T_p$ parameters
(Goroff et al. 1986; see Eq.~(\ref{SpTp}) below for their
definition). We use non-linear perturbation theory (PT) in  
order to provide a quantitative description of transients. 
For this purpose, we extend the standard formulation of PT to include
the full time-dependence of density and velocity solutions to arbitrary
non-linear order in the perturbation expansion. We show
that initial conditions can be thought as exciting non-linear decaying
modes in the evolution of perturbations which lead to transient
behavior. Although
we assume the initial conditions given by the ZA, the present
formalism can be useful to explore other non-Gaussian initial
conditions as well. We present results up to $p=8$, although the
techniques developed in this work make calculations possible to higher
orders if desired with minimal extra effort. Given that current angular surveys
are able to measure up to $S_9$ (Gazta\~naga 1994; Szapudi, Meiksin \&
Nichol 1996), and that the situation in redshift surveys will greatly
improve in the near future,  it is important to address the
issue of what requirements are needed in order to determine accurately these
statistical measures of clustering in numerical simulations.

The problem of transients from ZA initial conditions is well known
in the PT literature, and it has been pointed out many times before
(e.g., Juszkiewicz, Bouchet \& Colombi 1993, Juszkiewicz et al. 1995). 
However, the only quantitative understanding of the magnitude of this
problem so far comes from the numerical work of Baugh, Gazta\~naga \&
Efstathiou (1995, BGE hereafter), who studied the transients in the 
skewness of the
density field and found that for CDM simulations, an expansion $a \approx
3$ away from the initial conditions was
necessary to erase the memory of the ZA-induced skewness. 
Recently, the effect of transients
in $S_3$ at large scales was seen in the numerical renormalization
solution by Couchman \& Peebles (1997, Figure 12), where fluctuations
at large 
scales are applied by the ZA at each renormalization step and then
expanded by a factor of two in scale factor. For higher order
moments and velocity fields moments, there is no  analysis (numerical
or otherwise) available in the literature. In this sense, this work
constitutes  a natural development of the subject and should be useful
in designing numerical simulations of cosmological structure formation. 

% The use of PT has, as usual, its advantages and
% drawbacks when compared to an investigation via
% numerical simulations. The most important advantage is that the PT
% approach is analytical and therefore free of uncertainties that
% may render a numerical approach difficult, in the sense that other
% systematics such as finite volume effects play a role that can mimic the
% effects of transients (i.e. an underestimate of $S_p$ increasingly
% with $p$). This makes the PT approach very useful to distinguish the
% importance of both contributions and extract the dependencies on
% parameters of interest for numerical simulations. For example, as we
% shall see below, transients have an opposite dependence on spectral index
% from finite volume effects. The main drawback of the PT approach is
% that is only limited to the weakly non-linear regime, one would also like
% to investigate to what extent transients are a problem in the
% non-linear regime after shell-crossing has occurred. 

We now give a somewhat detailed summary of the contents of this paper
for those readers not familiar with the technical machinery of
non-linear PT and mostly interested in the relevance of the results
for numerical simulations. 

In Section~\ref{eom} we review the standard formulation of non-linear
PT. Eqs.~(\ref{pt_exp})-(\ref{pt_recrel}) show how to obtain non-linear
solutions to the equations of motion by a recursive procedure from
linear PT solutions, as first derived by Goroff et al. (1986). 

Section~\ref{za} presents the description of ZA initial
conditions in terms of Lagrangian trajectories
(as used to set initial conditions in numerical simulations, 
Eqs.~(\ref{Dza})-(\ref{Vza})) and in Eulerian space
(Eqs.~(\ref{zel-ansatz})-(\ref{ec:FZ})), 
which is in fact the most convenient way to calculate the statistical
properties of initial conditions. We explore two different ways of
setting initial velocities, the standard ZA approach which leads to
Eq.~(\ref{vel-corr}), and the EDFW scheme which yields
Eq.~(\ref{vel-corr2}) for the divergence of the initial peculiar
velocities.  

In Section~\ref{trans}, we extend the non-linear PT formalism
to include a description of transients. The main result in this regard is the
recursion relation in Eq.~(\ref{pt_recrel_sd}) which gives the
non-linear solutions to the equations of motion 
including the properties of
initial conditions (represented by the first term) and how they excite
non-linear decaying modes (transients). These solutions interpolate
between the initial conditions (ZA)  and the late-time 
exact-dynamics recursion relations given by Eq.~(\ref{pt_recrel}).

In Section~\ref{stat}, explicit expressions for the $S_p$ and $T_p$ are
given in terms of the solutions presented in the previous section. 
A simple way to understand the physical meaning of the $S_p$
parameters is provided by the 
Edgeworth expansion (e.g. Juszkiewicz et al. 1995; Bernardeau \&
Kofman 1995) for the probability distribution function (PDF) of the
standardized variable $\nu \equiv \d/\sigma$ 

\beq
\label{ew}
P(\nu)=P_{\rm G}(\nu)\ \Big\{1 + \frac{1}{6}\ S_3\ H_3(\nu)\ \sigma+
\Big[ \frac{S_4}{24}\ H_4(\nu) + \frac{S_3^2}{72}\ H_6(\nu) \Big]\
\sigma^2 + \ldots \Big\},
\eeq
where $P_{\rm G}(\nu) \equiv (2\pi)^{-1/2}\ \exp(-\nu^2/2)$ is the Gaussian
distribution, and $H_n(\nu)$ are the Hermite polynomials, e.g. 
$H_3(\nu)=\nu^3-3\nu$. An
equivalent expression can be written down for the velocity divergence PDF
involving the $T_p$ parameters. We see from Eq.~(\ref{ew}) that as
$\sigma \rightarrow 0$, the PDF approaches a Gaussian
distribution. The first-order correction to Gaussianity is given by the skewness
factor $S_3$ which, being proportional to the third moment of
the PDF (see Eq.~(\ref{SpTp})), measures the asymmetry between overdense and
underdense regions. Gravitational clustering leads to a
positive $S_3$ (Peebles 1980), which means that the high-density tail of the PDF 
is enhanced and the underdense tail is suppressed with respect to a 
Gaussian distribution, as expected from the attractive nature of the
gravitational force. Similarly, since underdense regions expand and 
occupy a larger fraction of the volume, the most probable value of $\d$ becomes 
negative, from Eq.~(\ref{ew}) the maximum of the PDF is in fact 
reached at  

\beq
\d_{{\rm max}} \approx -\frac{S_{3}}{2}\ \sigma,
\eeq 
to first order in $\sigma$. We thus see that the skewness factor 
$S_{3}$ contains very useful information on the shape of the PDF. 
Similarly, higher-order $S_{p}$ parameters provide further information 
on the PDF shape, for example, the next term in  Eq.~(\ref{ew}) contains a 
contribution from the kurtosis factor $S_{4}$ that gives the 
lowest-order measure of   flattening of the PDF tails relative to a 
Gaussian distribution. Comparison with exact PDF results when 
available (e.g. for the ZA where the PDF can be calculated 
non-perturbatively, see Bernardeau \& Kofman 1995) and numerical 
simulations (Juszkiewicz et al. 1995) shows that a few 
terms   in Eq.~(\ref{ew}) are good enough to 
reproduce the shape of the PDF near its maximum. On the other hand, 
the Edgeworth expansion rapidly breaks down at the tails of the 
distribution, even becoming non-positive definite, 
because they receive appreciable contributions 
from the whole hierarchy of $S_{p}$, and the Edgeworth series cannot 
be truncated at finite order. In fact, the use of powerful 
generating function techniques (Balian \& Schaeffer 1989; 
Bernardeau 1992, 1994; Bernardeau \& Kofman 1995), which can be 
thought as a resummation of the Edgeworth expansion, shows that the tails 
of the PDF in the asymptotic region are determined by the infinite 
series of $S_{p}$ parameters through their generating function. 

In summary, the main point of this digression is to convince the 
reader of the physical relevance of the $S_{p}$ and $T_{p}$ parameters as 
statistical measures of clustering. It should be clear from this 
discussion that an accurate determination of these is desired in 
numerical simulations in order to represent the statistics of 
large-scale structures correctly (BGE; Gazta\~naga \& Baugh 1995;
{\L}okas et al. 1995; Juszkiewicz et al. 1995; Colombi, Bouchet \&
Hernquist 1996; Bernardeau et al. 1997). 
Regarding the calculation of   $S_{p}$ and $T_{p}$   from 
leading-order (tree-level) PT  (that is, at lowest
non-vanishing order in the variance of the density field), 
it turns out that vertices
(defined in Eq.~(\ref{vert})) contain all the necessary information
about the non-linear dynamics. A recursion relation for vertices that
includes transients is presented in Appendix~\ref{appA}.

In Section~\ref{res} we present the perturbative  
results for transients in $S_{p}$ and $T_{p}$. Figure~\ref{fig1} 
shows the magnitude of transients for $S_{p}$ ($3\leq p\leq 8$) 
for different spectra, and similarly for $T_{p}$ in Fig.~\ref{fig2}. 
A comparison with numerical simulations with initial velocities as in 
EDFW is presented in Fig.~\ref{fig5}. Figures~\ref{fig11} 
and~\ref{fig12} present the corresponding transients in $S_{p}$ and 
$T_{p}$ when initial conditions are generated using second-order 
Lagrangian PT (2LPT). Note the significant improvement when compared 
to the equivalent plots in Figs.~\ref{fig1} and~\ref{fig2} for the 
ZA. 

Section~\ref{conc}
contains our conclusions and a final discussion. 
Technical material
regarding details of the calculations and further analytic results are
presented in Appendices~\ref{appA},~\ref{appB},
and~\ref{appC}. 
Finally, Appendix~\ref{appD} presents a simple
procedure to implement second-order Lagrangian PT initial conditions
in numerical simulations which requires
minimal numerically inexpensive modifications to a standard ZA code.

%
%
%=======================================================
\section{Dynamics}
\label{dyn}
%=======================================================
%
%

%
%
%=======================================================
\subsection{Standard Formulation of Perturbation Theory}
\label{eom}
%=======================================================
%
%
 
In the single-stream approximation, 
prior to orbit crossing, one can adopt a fluid description of the 
cosmological $N$-body problem, where  the relevant equations of motion 
correspond to conservation of 
mass and momentum and the Poisson equation (e.g., Peebles 1980).
Since vorticity decays in an expanding universe,
the system can be described completely in terms of the density 
contrast $\d (\x) \equiv [\rho(\x,t) - \bar \rho] / \bar \rho $ and
the velocity divergence, $ \theta \equiv \del \cdot \v $, where
$\v(\x)$ is the peculiar velocity.
Defining the conformal time $\tau=\int dt/a$, where 
$a(\tau)$ is the cosmic scale factor, and the conformal expansion
rate $ {\cal H}\equiv {d\ln a /{d\tau}}=H a$, where $H$ is the Hubble
constant, the equations of motion in
Fourier space become (e.g. Scoccimarro \& Frieman 1996)

\bml
\label{EOM}
\beqa
  &&{\partial \tilde{\delta}(\k,\tau) \over {\partial \tau}} +
	\tilde{\theta}(\k,\tau) = - \int d^3 k_1 \int d^3 k_2 
	\ \dD_2 \ \alpha(\k, \k_1) \ \tilde{\theta}(\k_1,\tau) \ 
	\tilde{\delta}(\k_2,\tau)   \label{ddtdelta}, \\
  &&{\partial \tilde{\theta}({\k},\tau) \over{\partial \tau}} +
	{\cal H}(\tau) \ \tilde{\theta}(\k,\tau) + {3 \over 2} \ \Omega 
	 \ {\cal H}^2(\tau) \ \tilde{\delta}(\k,\tau) =    
	- \int d^3k_1 \int d^3k_2 \ \dD_2 \ \beta(\k, \k_1, \k_2)
	\ \tilde{\theta}(\k_1,\tau) \ \tilde{\theta}(\k_2,\tau) ,\nonumber \\ 
	\label{ddttheta}
\eeqa
\eml 
where we used the Fourier transform representation $\dt(\k) = (2 
\pi)^{-3} \int d^3 x  \, \d(\x) \, e^{-i\ks\cdot\xs}$.  
In Eqs.~(\ref{EOM}),  $\k$ denotes a comoving wave number, $\dD_2 \equiv
\d_{\rm D}(\k-\k_1-\k_2)$, with $\d_{\rm D}$ the Dirac delta
distribution, and
\beq
	\alpha(\k, \k_1) \equiv {\k \cdot \k_1 \over{k_1^2}}, 
	\qquad \beta(\k, \k_1, \k_2) \equiv 
	{k^2 (\k_1 \cdot \k_2 )\over{2 \ k_1^2 \ k_2^2}}  \label{albe},
\eeq
which describe mode-coupling due to the non-linear dynamics.
Equations (\ref{ddtdelta}) and (\ref{ddttheta}) are valid in an
arbitrary homogeneous and isotropic universe, which evolves according
to the standard Friedmann equations.
In linear PT, the solution to the equations of motion 
(\ref{ddtdelta}) and (\ref{ddttheta}) is given by
\beqa
\tilde{\delta}(\k,\tau) &=& D(\tau) \ \delta_1(\k), \\
\tilde{\theta}(\k,\tau) &=& - {\cal H}(\tau) \ f(\Omega,\Lambda) \ D(\tau) \ 
\delta_1(\k),
\eeqa 
where $D(\tau)$ is linear growing mode, which from the
equations of motion must satisfy
\beq
\frac{d^2D}{d\tau^2} + {\cal H}(\tau) \ \frac{dD}{d\tau} =
\frac{3}{2} \ \Omega \ {\cal H}^2(\tau) \ D,
\label{D1}
\eeq
and $f(\Omega,\Lambda)$ is defined as 
\beq
f(\Omega,\Lambda) \equiv \frac{d \ln D}{d \ln a} = \frac{1}{\cal H}
\frac{d \ln D}{d \tau}.
\label{f}
\eeq
Explicit expressions or $D(\tau)$ and $f(\Omega,\Lambda)$ are
not needed for our purposes (see e.g. Peebles 1980), but useful  fits 
in the cosmologically interesting cases are 
$f(\Omega,\Lambda) \approx \Omega^{3/5}$ when $\Lambda=0$ (Peebles 1976), 
and  $f(\Omega,\Lambda) \approx \Omega^{5/9}$ when
$\Omega+\Omega_\Lambda=1$ (Bouchet et al. 1995). When $\Omega=1$, the 
linear growth factor $D(\tau)$ becomes the scale factor $a(\tau)$ and 
$f=1$. In this case, the PT solutions at each order $n$ scale as $a^n(\tau)$ 
and a general recursion relation is available that gives the PT solutions at 
arbitrary order (Goroff et al. 1986, see Eqs.~(\ref{pt_recrel}) below). 
When $\Omega \neq 1$ and/or $\Lambda \neq 0$, the PT solutions at each
order become increasingly more complicated, due to the fact that
growing modes at order $n$ in PT do not scale as $a^n(\tau)$.
Furthermore,
the solutions at each order become non-separable functions of $\tau$
and $\k$ (Bouchet et al. 1992, 1995; Bernardeau 1994b; Catelan et al. 
1995), and there appear to be no general recursion relations for the
PT kernels in an arbitrary FRW cosmology. However,  a
simple approximation to the equations of motion, $\Omega/f^2=1$, 
noted by Martel \& Freudling (1991) for second order PT, leads to 
separable solutions to arbitrary order in PT in which the linear
growth factor  
$D(\tau)$ plays the role of the scale factor, and the same recursion
relations as in the Einstein-de Sitter case are obtained for the PT 
kernels (Scoccimarro et al. 1998).
All the information on the
dependence of the PT solutions on the cosmological parameters $\Omega$
and $\Lambda$ is then encoded in the linear growth factor, 
$D(\tau)$, which in turn corresponds to the normalization of the linear
power spectrum. Equivalently,   in this case the equations of 
motion can  be written independently of cosmology by taking the linear 
growth factor as a time variable using Eq.~(\ref{f}) (Nusser \&
Colberg 1997).

The PT solutions for arbitrary cosmological models in this
approximation can be written down as (Scoccimarro et al. 1998,
Appendix B.3)

\bml
\label{pt_exp}
\beqa
\tilde{\delta}(\k,\tau) &=& \sum_{n=1}^{\infty} D^n(\tau) \ \delta_n(\k), \\
\tilde{\theta}(\k,\tau) &=& - {\cal H}(\tau) \ f(\Omega,\Lambda) 
\sum_{n=1}^{\infty} D^n(\tau) \ \theta_n(\k), 
\eeqa
\eml
The equations of motion  then determine $\delta_n(\k) $ and $\te_n(\k) $
in terms of the linear fluctuations,

\bml
\label{pt_exp2}
\beqa
	\delta_n(\k) &=& \int d^3k_1 \ldots \int d^3k_n \ 
    \dD_n \, F_n^{(s)} (\k_1, \ldots, \k_n)
	\  \delta_1(\k_1) \cdots \delta_1(\k_n)  \label{ec:deltan}, \\
	\theta_n({\k}) &=& \int d^3k_1 \ldots \int d^3k_n \ 
	\dD_n  \, G_n^{(s)}(\k_1, \ldots , \k_n) \ 
	\delta_1(\k_1) \ldots \delta_1(\k_n)   \label{ec:thetan}.
\eeqa
\eml
where $\dD_n \equiv \dD(\k- \k_1 - \cdots - \k_n)$, and the functions
$F_n^{(s)}$ and $G_n^{(s)}$ are constructed from the  
fundamental mode coupling functions $\alpha({\k}, {\k}_1)$ 
and $\beta({\k}, {\k}_1, {\k}_2)$ by a recursive procedure 
(Goroff et al. 1986; Jain \& Bertschinger 1994), 

\bml
\label{pt_recrel}
\beqa
F_n(\k_1,  \ldots , \k_n) &=& \sum_{m=1}^{n-1}
  { G_m(\k_1,  \ldots , \k_m) \over{(2n+3)(n-1)}} \Bigl[ (2n+1)\  
  \alpha(\k,\k^{(m)})\  F_{n-m}(\k_{m+1}, \ldots , \k_n) \nonumber \\
&& \qquad + 2\ \beta(\k,\k^{(m)}, \k^{(n-m)})\ 
  G_{n-m}(\k_{m+1}, \ldots , \k_n) \Bigr] \label{ec:Fn}, \\
G_n(\k_1,  \ldots , \k_n) &=& \sum_{m=1}^{n-1}
 { G_m(\k_1, \ldots , \k_m) \over{(2n+3)(n-1)}}
 \Bigl[3\ \alpha(\k,\k^{(m)})\  F_{n-m}(\k_{m+1}, \ldots , \k_n) \nonumber \\
&& \qquad + 2n\ \beta(\k, \k^{(m)}, \k^{(n-m)})\   
  G_{n-m}(\k_{m+1}, \ldots ,\k_n) \Bigr] \label{ec:Gn}
\eeqa
\eml
(where $ \k^{(m)} \equiv \k_1 + \ldots + \k_m$, 
$\k^{(n-m)} \equiv \k_{m+1} + \ldots + \k_n$,  
$\k \equiv \k^{(m)} +\k^{(n-m)} $, and $F_1 = G_1 \equiv
1$). Symmetrizing $F_n$ and $G_n$ over the wavectors $\k_1,  \ldots ,
\k_n$ leads to $F_n^{(s)}$ and $G_n^{(s)}$ as needed in Eq.~(\ref{pt_exp2}). 

%
%
%=======================================================
\subsection{Initial Conditions: Zel'dovich Approximation}
\label{za}
%=======================================================
%
%
 
Let us now discuss the perturbative formulation of the Zel'dovich
(1970) approximation, which we shall use to calculate the statistical
properties of the initial conditions. In the ZA, the motion of each 
particle $\x(\q,\tau$) at initial position $\q$ is obtained by applying
linear PT to its Lagrangian displacement, ${\bf \Psi}(\q,\tau)$, 

\beq
\label{Dza}
\x(\q,\tau) = \q + {\bf \Psi}(\q,\tau)  \approx \q - D(\tau) \ 
\del\phi^{(1)}(\q),
\eeq
where $\phi^{(1)}(\q)$ denotes a Lagrangian potential given by the
initial conditions, see 
Appendix~\ref{appD} for a more detailed discussion of ZA in Lagrangian
space. This 
implies that the velocities of particles initially at $\q$ are given by

\beq
\label{Vza}
\v \approx - D(\tau)\ {\cal H}(\tau)\ f \ \del\phi^{(1)}(\q).
\eeq
In Eulerian space, it can be shown that the ZA is equivalent to the
dynamics obtained by replacing the Poisson equation by the  
ansatz (Munshi \& Starobinsky 1994, Hui \& Bertschinger 1996)

\begin{equation}
	\v(\x,\tau) = -\frac{2 f}{3 \Omega {\cal H}(\tau)} \del 
	\Phi(\x,\tau)
	\label{zel-ansatz},
\end{equation}
which is the relation between velocity $\v(\x,\tau)$ and
gravitational potential $\Phi(\x,\tau)$ valid in linear PT 
theory. The ZA therefore fails to conserve momentum, i.e. the Euler
equation will only be satisfied to linear order. When calculating
one-point cumulants  of density and velocity fields, is
convenient to use the perturbative version of ZA, since the
non-perturbative probability distribution function (PDF) is singular
due to caustic formation in the 
sense that  
one-point cumulants $\lexp \d^p \rexpc$ of the density field diverge 
for $p>1$ due to the high-density tail (e.g., see Bernardeau \& Kofman
1995). The perturbative 
formulation of ZA follows from replacing Eq.~(\ref{zel-ansatz}) into
the equations of motion~(\ref{EOM}), which yields the corresponding recursion
relations (Scoccimarro \& Frieman 1996) 

\bml
\label{ZA_recrel}
\beqa
F_n^{\rm ZA}(\k_1,  \ldots , \k_n) &=& \sum_{m=1}^{n-1}
  { G_m^{\rm ZA}(\k_1,  \ldots , \k_m) \over{n(n-1)}} \Bigl[ (n-1)\  
  \alpha(\k,\k^{(m)})\  F_{n-m}^{\rm ZA}(\k_{m+1}, \ldots , \k_n) \nonumber \\
&& \qquad \qquad \qquad \qquad \qquad +  \beta(\k,\k^{(m)}, \k^{(n-m)})\ 
  G_{n-m}^{\rm ZA}(\k_{m+1}, \ldots , \k_n) \Bigr] \label{ec:FnZA}, \\
G_n^{\rm ZA}(\k_1,  \ldots , \k_n) &=& \frac{1}{(n-1)}\sum_{m=1}^{n-1}
 G_m^{\rm ZA}(\k_1, \ldots , \k_m)\ \beta(\k, \k^{(m)}, \k^{(n-m)})\   
  G_{n-m}^{\rm ZA}(\k_{m+1}, \ldots ,\k_n)  \label{ec:GnZA}.
\eeqa
\eml
Upon symmetrization these give a
simple result for the density field PT kernels (Grinstein \& Wise 1987)

\beq
 F_n^{\rm ZA}(\k_1,  \ldots  ,\k_n) = \frac{1}{n!}
 \frac{({\k} \cdot \k_1)}{k_1^2} \ldots 
 \frac{({\k} \cdot \k_n)}{k_n^2} 
 \label{ec:FZ}.
\eeq
In numerical simulations, the ZA is generally used to set the initial
perturbations as follows (see e.g. EDFW). 
A Gaussian density field, $\d_\ell(\k)$ is generated in Fourier space
from the desired linear power spectrum, and therefore the Lagrangian
potential $\tilde\phi^{(1)}(\k)=-\d_\ell(\k)/k^2$ is Fourier
transformed and its gradient taken to yield $\del_q \phi^{(1)}(\q)$ at
each ``unperturbed'' particle position, denoted by $\q$, which is
usually described by a grid or a ``glass''. This gives the 
displacement field which is used in Equation~(\ref{Dza}) to
displace the particles from their unperturbed positions and imposes
the ZA density 
field from the initial Gaussian field, i.e. $\delta^{\rm ZA}(\x)= {\cal M}
[\delta_\ell(\q)]$, where ${\cal M}$ denotes the ZA mapping implied by
Eqs.~(\ref{ec:deltan}) and (\ref{ec:FZ}). Equation~(\ref{Vza})
can then be used to assign the velocities
to particles. In this case, the velocities then satisfy

\beq
\theta(\x) = -f \ {\cal H} \
\sum_{n=1}^{\infty} D^n(\tau) \  \int d^3k_1 \ldots \int d^3k_n \ 
	\dD_n  \, G_n^{\rm ZA}(\k_1, \ldots , \k_n) \ 
	\delta_\ell(\k_1) \ldots \delta_\ell(\k_n).
\label{vel-corr}
\eeq
It is in fact straightforward but tedious to show that Eq.~(\ref{Vza})
in Lagrangian space implies Eq.~(\ref{vel-corr}) in Eulerian space
with the kernels $G_n^{\rm ZA}$ given by Eq.~(\ref{ec:GnZA}). 

Although most existing initial conditions codes use this prescription
to set up their 
ZA initial conditions, there is another prescription to set initial
velocities suggested by EDFW, which avoids the high initial
velocities that result from Eq.~(\ref{Vza}) because of 
small-scale density fluctuations approaching unity when starting a
simulation at late epochs (low redshift). This procedure corresponds to  
recalculate the velocities from the gravitational potential due to 
the perturbed particle positions, obtained by solving again Poisson
equation after particles have been displaced according to
Eq.~(\ref{Dza}). Linear PT is then applied to the density field to
obtain the velocities, which implies instead of Eq.~(\ref{Vza}) and
Eq.~(\ref{vel-corr}) that  

\beq
\theta(\x) = -f \ {\cal H} \ \delta^{\rm ZA}(\x)= -f \ {\cal H} \
\sum_{n=1}^{\infty} D^n(\tau) \  \int d^3k_1 \ldots \int d^3k_n \ 
	\dD_n  \, F_n^{\rm ZA}(\k_1, \ldots , \k_n) \ 
	\delta_\ell(\k_1) \ldots \delta_\ell(\k_n).
\label{vel-corr2}
\eeq
Therefore, in this case, {\em the initial velocity
field is such that the divergence field $\Theta(x) \equiv
\theta(\x)/(-f \ {\cal H})$ has the same higher-order correlations as
the ZA density perturbations}. Comparing Eq.~(\ref{vel-corr}) and
(\ref{vel-corr2}) we see that both prescriptions
agree in linear PT, where $F_1^{\rm ZA}=G_1^{\rm ZA}=1$, but they
differ at second and higher-order PT, since $F_n^{\rm ZA}\neq G_n^{\rm
ZA}$ for $n>1$. This implies that these two different alternatives of
setting the initial velocities affect the magnitude of the
transients, as we shall see below. It turns out that in fact, the
prescription given by Eq.~(\ref{vel-corr2}) is closer to the exact
dynamics given by Eqs.~(\ref{ec:thetan}) and~(\ref{ec:Gn}) than
Eq.~(\ref{vel-corr}), therefore, 
the second method will excite less decaying modes and
consequently transients will decay faster than in the
standard ZA scheme.

%
%
%=======================================================
\subsection{Transients in Perturbation Theory}
\label{trans}
%=======================================================
%
%

Perturbation theory describes the non-linear dynamics as a collection
of linear waves, $\d_1(\k)$, interacting through the mode-coupling
functions  $\alpha$
and $\beta$ in Eq.~(\ref{albe}). Even if the initial conditions
are set in the growing mode, after scattering due to  non-linear interactions
 waves do not remain purely in the growing mode. 
In the standard treatment, presented in Section~\ref{eom}, the 
subdominant time-dependencies that necessarily appear due to this 
process  have been neglected, i.e., only the 
fastest growing mode (proportional to $D^{n}$) 
is taken into account at each order $n$ in PT. In
this subsection we generalize the previous results to include the full
time dependence of the solutions at every order in PT. This is 
necessary to properly address the problem of transients from ZA
initial conditions.

In this case we write the perturbative solutions as (see
Eqs.~(\ref{pt_exp})): 

\bml 
\label{ptexp}
\beqa
\tilde{\delta}(\k,\tau) &=& \sum_{n=1}^{\infty} D^n(\tau) \ \delta_n(\k,D), \\
\tilde{\theta}(\k,\tau) &\equiv & - {\cal H}(\tau) \ f(\Omega,\Lambda)  
\ \Theta(\k,\tau) = 
- {\cal H}(\tau) \ f(\Omega,\Lambda) \sum_{n=1}^{\infty} D^n(\tau) \ 
\theta_n(\k,D), \label{Theta} 
\eeqa
\eml
where $\d_n$ and $\te_n$ are written in terms of PT kernels as in
Equation~(\ref{pt_exp2})

\beq
\label{pt_exp3}
	\Phi_a^{(n)}(\k,D) = \int d^3k_1 \ldots \int d^3k_n \  
\dD_n \ {\cal F}_a^{(n)} (\k_1, \ldots, \k_n;D)
	\ \delta_1(\k_1) \cdots \delta_1(\k_n)  \label{Phi_n},
\eeq
where $a=1,2$ and $\Phi_1^{(n)} \equiv \d_n$, $\Phi_2^{(n)} \equiv
\te_n$. The kernels ${\cal F}_a^{(n)}$ now depend on the linear growth
factor $D(\tau)$ and reduce to the ones in the previous 
section when transients die
out, that is ${\cal F}_1^{(n)}\rightarrow F_n$, ${\cal
F}_2^{(n)}\rightarrow G_n$ when $D(\tau) \rightarrow \infty$. Also,
Eq.~(\ref{pt_exp3}) incorporates in a convenient way initial
conditions, if we assume that at $D=1$, ${\cal F}_a^{(n)} = {\cal
I}_a^{(n)}$, {\em we effectively reduce the non-Gaussianity of initial
conditions to a Gaussian problem, i.e. where the linear solutions
$\d_1(\k)$ are Gaussian random fields}. The kernels ${\cal I}_a^{(n)}$
describe the initial correlations imposed at the start of the
simulation, for the standard ZA scheme we have (see Eq.~(\ref{vel-corr}))

\beq
{\cal I}_1^{(n)}=F_n^{\rm ZA}, \ \ \ \ \ {\cal I}_2^{(n)}=G_n^{\rm ZA},
\eeq
whereas for velocities set from perturbed particle positions we have
instead (see Eq.~(\ref{vel-corr2}))

\beq
{\cal I}_1^{(n)}=F_n^{\rm ZA}, \ \ \ \ \ {\cal I}_2^{(n)}=F_n^{\rm ZA}.
\eeq
The
recursion relations for ${\cal F}_a^{(n)}$, which solve the non-linear
dynamics at arbitrary order in PT, can be obtained by replacing
Eq.~(\ref{Phi_n}) into the equations of motion, which yields   

\beqa
\label{pt_recrel_sd}
{\cal F}_a^{(n)}(\k_1,  \ldots , \k_n;z) &=& 
{\rm e}^{-n z} \ g_{ab}(z) \ {\cal I}_b^{(n)}(\k_1,  \ldots , \k_n) + 
\sum_{m=1}^{n-1}
\int_0^z ds  \ {\rm e}^{n(s-z)} \ g_{ab}(z-s) \ 
\gamma_{bcd}(\k,\k^{(m)},\k^{(n-m)}) 
\nonumber \\& &   \times \ {\cal F}_c^{(m)}(\k_1,  \ldots , 
\k_m;s) \ {\cal F}_d^{(n-m)}(\k_{m+1}, \ldots , \k_n;s),
\eeqa
where $z \equiv \ln D(\tau)$ and we have assumed the
summation convention over repeated indices, which run between 1 and
2. In Equation~(\ref{pt_recrel_sd}) the matrix

\beq
g_{ab}(z) = \frac{{\rm e}^z}{5}
\Bigg[ \begin{array}{rr} 3 & 2 \\ 3 & 2 \end{array} \Bigg] -
\frac{{\rm e}^{-3z/2}}{5}
\Bigg[ \begin{array}{rr} -2 & 2 \\ 3 & -3 \end{array} \Bigg],
\label{prop}
\eeq
is the {\em linear propagator} (Scoccimarro 1997b). The first term 
represents the propagation of linear growing mode
solutions, where the second corresponds to the decaying modes
propagation. That is, in linear perturbation theory, density and velocity
perturbations propagate in time according to

\beq
\Phi_a^{(1)}(z) = g_{ab}(z) \ \Phi_b^{(1)}(0).
\eeq
If initial conditions are set in the growing mode, then
$\Phi_a^{(1)}(0) \propto (1,1)$ vanishes upon contraction with the
second term in  $g_{ab}(z)$, whereas the first term reduces to the
familiar $\Phi_a^{(1)}(z) = {\rm e}^z \Phi_a^{(1)}(0) = D(\tau)\ 
\Phi_a^{(1)}(0)$.  The 
``scattering matrix'' $\gamma_{abc}$ encodes the 
non-linear interactions and is given by

\bml
\label{scatt-m}
\beqa
\gamma_{121}(\k,\k_1,\k_2)&=&\alpha(\k,\k_1), \\
\gamma_{222}(\k,\k_1,\k_2)&=&\beta(\k,\k_1,\k_2),
\eeqa
\eml
and zero otherwise, with $\k \equiv \k_{1}+\k_{2}$.
Note that since $g_{ab}(z) \rightarrow \delta_{ab}$ as $z \rightarrow 
0^{+}$, 
in Eq.~(\ref{pt_recrel_sd}) the kernels 
${\cal F}_a^{(n)}$ reduce to ${\cal I}_a^{(n)}$  at $D=1$, where the
initial conditions are 
set. Equation~(\ref{pt_recrel_sd}) reduces
to the standard recursion relations in Eq.~(\ref{pt_recrel}) for 
Gaussian initial conditions (${\cal I}_a^{(n)}=0$ for $n>1$) when
transients are neglected, i.e. the time dependence of ${\cal F}_a^{(n)}$ 
is neglected and the lower limit of integration is
replaced by $s=-\infty$. 
Also, it is easy to check from Eq.~(\ref{pt_recrel_sd}) that if 
${\cal I}_a^{(n)}  = (F_{n},G_{n}) $,  then ${\cal F}_a^{(n)} = 
(F_{n},G_{n})$, as it should be. 
In what follows, the kernels in
Eq.~(\ref{pt_recrel_sd}) are assumed to be symmetrized over its
arguments. Note that these kernels
are no longer a separable function of wave-vectors and time. 

Equation~(\ref{pt_recrel_sd}) gives useful insight 
into the nature of nonlinear PT solutions. For example, second order
solutions are  
built from the interaction (represented by the $\gamma$ matrix) of two 
linear waves that are propagated freely to the present time using 
Eq.~(\ref{prop}), plus the contribution from initial conditions 
propagated to the present represented by the first term in 
Eq.~(\ref{pt_recrel_sd}). 
After scatterings, waves do not remain purely in 
their growing modes, and from Eq.~(\ref{pt_recrel_sd}) one can check
that there is a contribution from decaying modes 
propagation even for the fastest growing mode included in standard 
treatments of PT. More importantly for the present purposes is the 
fact that the contribution to the PT solutions at a given order $n$ 
and ``time'' $z$ depend on all the $n$-scattering processes that 
happened between $s=z$ and $s=1$, where initial conditions are set. 
By assuming that initial conditions are set in the ``infinite past'' 
($s=-\infty$), the standard formulation of PT presented in the 
previous section contains no information on the time scale that takes 
to erase the memory of the correlations imposed by the initial 
conditions. A more detailed treatment of this formalism and 
other applications is left for a future paper (Scoccimarro 1997b).

%
%
%=======================================================
\section{Statistics}
\label{stat}
%=======================================================
%
%

We are interested in one-point cumulant statistics for the density and
velocity divergence fields, which can be
characterized by the $S_p$ and $T_p$ parameters defined as

\beq
\label{SpTp}
S_p(R) \equiv \frac{\lexp \d^p(R) \rexpc}{\lexp \d^2(R) \rexp^{p-1}}
\ \ \ \ \ \ \ \ \ \ 
T_p(R) \equiv \frac{\lexp \Theta^p(R) \rexpc}{\lexp \Theta^2(R) \rexp^{p-1}},
\eeq
where $R$ is the smoothing scale, and the subscript $c$ denotes
the connected contribution, i.e. the gaussian value is subtracted
off. Note that we define cumulants for 
the velocity divergence field in terms of $\Theta(\x)$ as defined in 
Eq.~(\ref{Theta}), which differs from the standard convention 
by a factor $(-1)^{p}$. In this work, we use top-hat
smoothing, which is described by a window function $W_{\rm TH}(k R)$

\beq
W_{\rm TH}(x)=\sqrt{\frac{9\pi}{2}} \ \frac{J_{3/2}(x)}{x^{3/2}} = 
\frac{3}{x^3} \ [\sin (x) -x \cos (x)]
\eeq
in the Fourier domain, where $J_\nu(x)$ is a Bessel function. To simplify the
notation, we henceforth denote:

\beq
d\sigma^2_i \equiv 4 \pi \ k_{i}^{2} \ dk_i \ P(k_i) \ \ \ \ \ \ \ \ \ \ 
W_{i \ldots j} \equiv W_{\rm TH}(|\k_i+ \ldots + \k_j| R),
\eeq
where $P(k)$ is the power spectrum, and $\sigma^2$ denotes the
variance of the density field fluctuations

\beq
\lexp \d (\k) \d (\k') \rexp \equiv \d_{\rm D}(\k+\k') \ P(k)
\ \ \ \ \ \ \ \ \ \ 
\sigma^2(R) \equiv \int d^3k \ P(k) \ W^2_{\rm TH}(kR).
\eeq

A systematic framework for  calculating  correlations of cosmological
fields in PT has been formulated using
diagrammatic techniques (Fry 1984; Goroff et al.~1986; Wise 1988;
Scoccimarro \& Frieman 1996). From this point of view, leading order
PT for the statistical quantities of interest
corresponds to tree graphs, next-to-leading order PT contributions
can be described in terms of one-loop graphs, etc. These diagrammatic
techniques assume Gaussian initial conditions, although in principle
they can be extended to any non-Gaussian model by adding the appropriate
new vertices (Wise 1988). In this work we are interested in a very particular
non-Gaussian initial condition, that given by the ZA as usually 
imposed to start numerical simulations. As shown in the previous
section, in this case one can include the non-Gaussianity directly
into the non-linear solutions and therefore the problem reduces
effectively to one dealing with Gaussian initial conditions.

We now write down the standard expressions for the $S_p$ parameters in
terms of non-linear kernels. In the discussion that follows, analogous analysis
always applies to the velocity divergence. In the
following we shall use the scale factor $a(\tau)$ to denote the time
dependence, {\em but in models with density parameter $\Omega \neq 1$ the
same equations will be valid upon replacing $a(\tau)$ by the linear
growth factor $D(\tau)$}. For the skewness factor, we have 

\beq
S_3(R,a) = 
\frac{6}{\sigma^4(R)} \int P_1 d^3 k_1 \ P_2 d^3 k_2 \  
W_1  W_{12} W_2  \ {\cal F}_1^{(2)}(\k_1,\k_2;a). 
\eeq
The calculation of tree-level diagrams such as this and the
expressions that follow is simplified by using the fact that tree-level PT
corresponds to taking the spherical average of PT kernels (Bernardeau
1992, 1994b). We therefore define  the angular
averaged smoothed kernels $\omega_a^{(n)}$ by ($a=1,2$):

\beq
\label{vert}
\frac{1}{n!}\ (\nu_n, \mu_n) \equiv \omega_{a}^{(n)} \equiv 
\int \ \Big(\frac{d\Omega_{1}}{4\pi}\Big) \ldots 
\Big(\frac{d\Omega_{n}}{4\pi}\Big) \ W_{1\ldots n} \ {\cal 
F}_{a}^{(n)}(\k_{1},\ldots,\k_{n}),
\eeq
whose recursion relations for top-hat smoothing are obtained directly from 
Eqs.~(\ref{pt_recrel_sd}) in Appendix A. The vertices $\nu_n$ and $\mu_n$
are the (Eulerian) smoothed counterpart of those defined in Bernardeau
(1992). Recently, Lagrangian smoothed vertices have been defined by
Fosalba \& Gazta\~naga (1997). The skewness factor smoothed at scale $R$ can then
be rewritten as:

\beq
S_3(R,a) = 
\int \frac{d\sigma^2_1 \ d\sigma^2_2}{\sigma^4(R)} \  
3 \ \nu_1(x_1;a)  \ \nu_2(x_1,x_2;a) \ \nu_1(x_2;a)  , 
\eeq
where $x_i \equiv k_i R$ and $\nu_1(x)=W(kR)$. To simplify the
notation, we henceforth suppress the dependence on the scale factor
$a$. Similarly, the final expression for the kurtosis factor is
 
\beqa
S_4(R) &=& \int \frac{d\sigma^2_1 \ d\sigma^2_2 \
d\sigma^2_3}{\sigma^6(R)} \ \Big[ 12 \ 
\nu_1(x_1)\ \nu_2(x_1,x_2) \ \nu_2(x_2,x_3)
\ \nu_1(x_3)+ 4 \ \nu_1(x_1)\ \nu_1(x_2)\
\nu_1(x_3)\   \nu_3(x_1,x_2,x_3) \Big],
\nonumber \\& &
\label{S4res}
\eeqa
whereas for $S_5$ (the ``pentosis'' parameter
according to the nomenclature of Chodorowski \& Bouchet 1996), we
obtain  
 
\beqa
S_5(R) &=&  \int \frac{d\sigma^2_1 \ d\sigma^2_2 \
d\sigma^2_3 \ d\sigma^2_4}{\sigma^8(R)} \  \Big[ 
60 \ \nu_1(x_1)\ \nu_2(x_1,x_2) \ \nu_2(x_2,x_3) \ 
\nu_2(x_3,x_4) \ \nu_1(x_4) + \ 60 \ \nu_1(x_1)\ \nu_1(x_2) 
\nonumber \\
& & \times  \ \nu_3(x_1,x_2,x_3)\ \nu_2(x_3,x_4) \ \nu_1(x_4) + 5 
\  \nu_1(x_1)\ \nu_1(x_2)\
\nu_1(x_3)\   \nu_1(x_4)\ \nu_4(x_1,x_2,x_3,x_4) \Big],
\label{S5res}
\eeqa
and, finally, for $S_6$ we have

\beqa
S_6(R) &=& \int \frac{d\sigma^2_1 \ d\sigma^2_2 \
d\sigma^2_3 \ d\sigma^2_4 \  d\sigma^2_5}{\sigma^{10}(R)} \ \Big[6 \ 
\nu_1(x_1)\ \nu_1(x_2)\ \nu_1(x_3)\ \nu_1(x_4)\ \nu_1(x_5)\
\nu_5(x_1,x_2,x_3,x_4,x_5) 
\nonumber \\
& & + \ 120 \ \nu_1(x_1)\ \nu_1(x_2)\ \nu_1(x_3)\ 
\nu_4(x_1,x_2,x_3,x_4)\ \nu_2(x_4,x_5)\ \nu_1(x_5)\nonumber \\
& & + \ 90 \  \nu_1(x_1)\ \nu_1(x_2)\  
\nu_3(x_1,x_2,x_3)\ \nu_3(x_3,x_4,x_5)\ \nu_1(x_4)\ \nu_1(x_5) \nonumber \\
& & + \ 360 \ \nu_1(x_1)\ \nu_1(x_2)\ 
\nu_3(x_1,x_2,x_3)\ \nu_2(x_3,x_4)\ 
\nu_2(x_4,x_5)\ \nu_1(x_5) \nonumber \\
& & + \ 360 \ \nu_1(x_1)\  
\nu_3(x_1,x_2,x_3)\ \nu_2(x_2,x_4)\ 
\nu_2(x_3,x_5)\ \nu_1(x_4)\ \nu_1(x_5) \nonumber \\
& & + \ 360 \ \nu_1(x_1)\ 
\nu_2(x_1,x_2)\ \nu_2(x_2,x_3)\ 
\nu_2(x_3,x_4)\ \nu_2(x_4,x_5)\ \nu_1(x_5)
\Big],
\label{S6res}
\eeqa

The tree-like structure of these expressions is quite obvious.  
The combinatorial factors in each term corresponds to the number of
labellings of each particular tree diagram; see e.g Fry (1984) for
$3\leq p \leq 6$ and Bosch\'an, Szapudi \& Szalay (1994) for up to
$p=8$.  The expressions for $S_{7}(R)$ and $S_{8}(R)$ will not be
reproduced here, but  
note that they can easily be obtained from the tree diagrams together with 
their combinatorial coefficients shown in table 1 of Bosch\'an, Szapudi 
\& Szalay (1994). The equivalent expressions for $T_p$ are generated
by simply replacing $\nu$ by $\mu$ in the formulas above.

%
%
%======================================================
\section{Results}
\label{res}
%======================================================
%
%

Despite the
complicated appearance of the  expressions given in the previous 
section, they can be 
calculated in a straightforward manner thanks to the special
geometrical properties of top-hat 
smoothing (Bernardeau 1994, see also Hivon et al. 1995). Gaussian
smoothing does not share these properties and will not be considered
here (see {\L}okas et al. 1995). The recursion relations for the 
smoothed vertices $\nu_{n}$ and $\mu_{n}$ as a function of scale
factor  $a$ and smoothing scale $R$ are derived in Appendix A from the
recursion relations given by Eq.~(\ref{pt_recrel_sd}). These vertices
depend on scale $R$  
through derivatives of the window functions, which are then 
converted into derivatives of the variance with 
respect to scale, the $\gamma_{p}(R)$ parameters defined by (Bernardeau 1994)

\beq
\label{gamma_p}
\gamma_{p}(R) \equiv - \frac{d^p \ln \sigma^{2}(R)}{d \ln^p R}.
\eeq 
Using the results in Appendix A, we get for the skewness parameters
($\gamma \equiv \gamma_1 = n_{\rm eff}+3$)

\bml
\label{p=3za}
\beqa
S_3(a) &=& \frac{1}{a} [4- \gamma] + \Big\{ \frac{34}{7} - \gamma \Big\}
+ \frac{1}{a}\Big(\gamma -\frac{26}{5} \Big) + \frac{12}{35a^{7/2}}
= \frac{34}{7} - \gamma - \frac{6}{5 a}+ \frac{12}{35 a^{7/2}},
\label{S3g} \\ 
& & \nonumber \\ 
T_3(a) &=& \frac{1}{a} [2- \gamma] + \Big\{ \frac{26}{7} - \gamma \Big\}
+ \frac{1}{a}\Big(\gamma -\frac{16}{5} \Big) - \frac{18}{35 a^{7/2}}
= \frac{26}{7} - \gamma - \frac{6}{5 a}- \frac{18}{35 a^{7/2}},
\label{T3g}
\eeqa
\eml
where we have assumed ZA initial velocities,
Eq.~(\ref{zel-ansatz}). On the other hand,  for  initial velocities
set from perturbed particle positions, as in 
Eq.~(\ref{vel-corr2}), we have:

\bml
\label{p=3}
\beqa
S_3(a) &=& \frac{1}{a} [4- \gamma] + \Big\{ \frac{34}{7} - \gamma \Big\}
+ \frac{1}{a}\Big(\gamma-\frac{22}{5} \Big) - \frac{16}{35 a^{7/2}}
= \frac{34}{7} - \gamma - \frac{2}{5 a}- \frac{16}{35 a^{7/2}} ,
\label{S3} \\ 
& & \nonumber \\ 
T_3(a) &=& \frac{1}{a} [4- \gamma] + \Big\{ \frac{26}{7} - \gamma
\Big\}+ \frac{1}{a}\Big(\gamma-\frac{22}{5} \Big)
+  \frac{24}{35 a^{7/2}} = \frac{26}{7} - \gamma - \frac{2}{5 a}
+ \frac{24}{35 a^{7/2}}.
\label{T3}
\eeqa
\eml
{\em For $\Omega \neq 1$, these expressions and the ones that 
follow are valid upon replacing the scale factor $a$ by the linear 
growth factor $D(\tau)$}. For scale-free initial conditions, with 
spectral index $n$, the results for top-hat smoothing are restricted 
to $n<1$, since for $n=1$ the variance in top-hat spheres diverges 
and the logarithmic derivative $\gamma$ becomes meaningless (the same 
restriction applies to $p>3$ results). The first term in square brackets in 
Eqs.~(\ref{p=3za}) and~(\ref{p=3}) represents 
the initial skewness given by the ZA (Bernardeau 1994), which decays 
with the expansion as $a^{-1}$ (Fry \& Scherrer 1994). The  second
and remaining terms in Eqs.~(\ref{p=3za}) and~(\ref{p=3}) represent the
asymptotic exact values (in between  
braces; Juszkiewicz, Bouchet \& Colombi 1993, Bernardeau 1994) and the
transient induced by the exact dynamics  respectively; 
their sum vanishes at $a=1$ where the only 
correlations are those imposed by the initial conditions. Similar
results to these are obtained for higher-order moments, we refer the
reader to  Appendix B for explicit expressions. Note that for
scale-free initial conditions, the transient contributions to $S_p$
and $T_p$ break self-similarity. 

\begin{figure}[t!]
\centering
\centerline{\epsfxsize=18truecm\epsfysize=18truecm\epsfbox{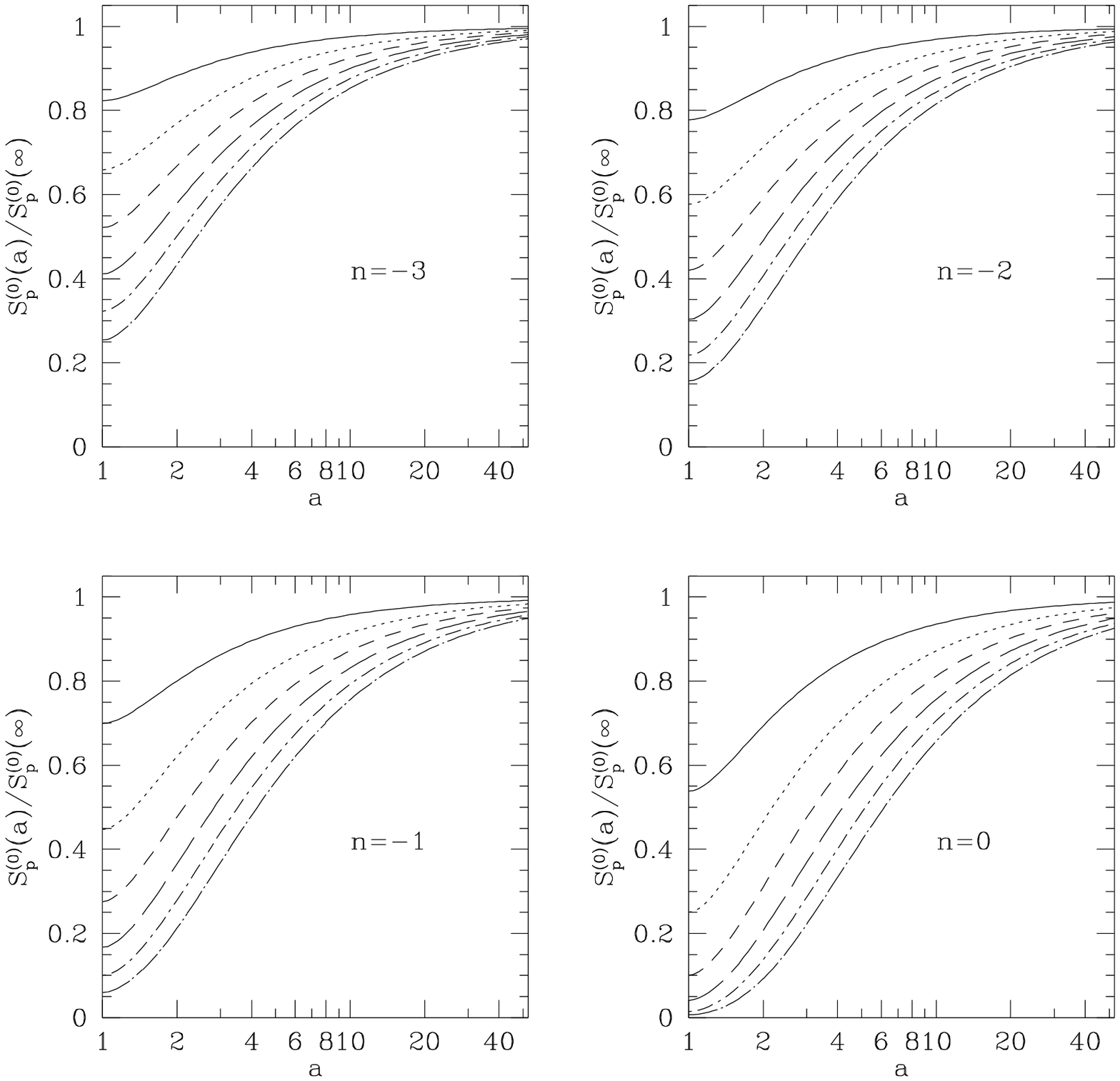}}
\caption{
The ratio of the tree-level $S_p$ parameters at scale factor
$a$ to their  asymptotic exact dynamics value for scale-free
initial spectra with different spectral indices: $S_3$ (solid lines), 
$S_4$ (dotted), $S_5$ (short-dashed), $S_6$ (long-dashed), $S_7$
(dot-short-dashed), and $S_8$ (dot-long-dashed). 
The values at 
$a=1$ represent those set by the ZA initial conditions. 
For cosmological models with $\Omega < 1$, the
horizontal axis becomes the linear growth factor.} 
\label{fig1}
\end{figure}

Figure~\ref{fig1} shows these results as a function of scale factor
$a$ for different spectral indices, assuming that velocities are set
as in the ZA. The plots show the
ratio of $S_p(a)$ to its ``true'' asymptotic value predicted by
PT, $S_p(\infty)$, for $3 \leq p
\leq 8$. The values at $a=1$ correspond to the ratio of ZA to exact
dynamics $S_p$'s, which becomes smaller as either $p$ or $n$
increases. For the skewness, it takes as much as $a=6$ for $n=0$ to
achieve 10\% of the asymptotic exact PT value, whereas spectra
with more  
large-scale power, where the ZA works better, require less expansion 
factors to yield the same accuracy. A similar
constraint should hold for the bispectrum, the three-point function
in Fourier space (Peebles 1980; Fry, Melott
\& Shandarin 1993; Scoccimarro et al. 1998). As $p$ increases,
however, the transients become worse and at $p=8$ an expansion by a
factor $a=40$ is required for $n=0$ to achieve $10\% $ accuracy in
$S_{8}$. This suggests that the tails of the PDF could be quite affected by
transients from initial conditions. Furthermore, for models where
$\Omega < 1$, the requirements 
on $a$ translate into requirements on the linear growth factor
$D(\tau)$, which implies a more stringent constraint on $a$, i.e. an
$\Omega < 1$ simulation should be started earlier (at a higher
redshift) than an $\Omega = 1$ model. For example, an open model
with $\Omega = 0.3$ typically requires a factor of two higher
initial redshift than for $\Omega = 1$ (see Figs.~\ref{fig7} and
~\ref{fig8} below).

Figure~\ref{fig2} shows the corresponding results for the velocity
divergence $T_p$ parameters. We see that the effects of transients in
this case is more severe than in the density field case, in particular
as $n$ increases, since the initial $a=1$ values imposed by the
initial conditions become quite different from the asymptotic exact PT
values. For example, $n=0$ requires $a \approx 15$ for $10\% $
accuracy in $T_3$ (more than a factor of two larger than for $S_3$). The
situation quickly deteriorates as $p$ increases. Again, the shape of
the PDF of the velocity divergence should be quite sensitive to the
presence of transients. Moreover, since statistics of the density
field in redshift space contain contributions from velocities
correlations, one expects that the redshift-space density
field PDF will be more affected by transients than in real space.

Figures~\ref{fig3} and~\ref{fig4} show the equivalent results for
initial velocities set from perturbed particle positions, 
Eq.~(\ref{vel-corr2}). Comparing to the corresponding results in
Figs.~\ref{fig1} and~\ref{fig2}, these results show considerable
improvement on the amount of expansion required to erase transients,
by factors of two or three depending on the spectrum.  
We see that at most $a \approx 3$ is necessary to 
achieve $10\% $ accuracy in $S_{3}$, at least a factor two better than
in the ZA velocities case.  These results are in agreement 
with the numerical study of BGE for CDM models, with velocities
assigned as in Eq.~(\ref{vel-corr2}),
in which they found that $a \approx 3$ was needed to 
recover the PT prediction for $S_{3}$ at scales where $n_{\rm eff}
\approx -1$. 

% Excitation of decaying modes in second and
% higher order PT causes the transients to take longer than in the
% previous case, although as $n$ increases the difference is not as
% large as for the $n=-3$ case. The results for Gaussian initial
% velocities are weakly sensitive to spectral index, which can be seen
% from Eqs.~(\ref{p=3za}) by taking the ratio to the asymptotic values.
 
Figure~\ref{fig5} presents a comparison of the perturbative
predictions for transients in $S_p$ parameters with the standard CDM
numerical simulations measurements 
of BGE, kindly
provided by E.~Gazta\~naga. The simulation evolved 
$100^3$ particles in a box 300 h$^{-1}$ Mpc a side. Initial velocities
are set according to Eq.~(\ref{vel-corr2}) as described in EDFW. The
error bars in 
the measurements correspond to the variance over 10 realizations. 
We plot these N-body results by taking the ratio to the tree-level
exact dynamics value predicted by PT, which has the advantage of
reducing the main scale dependence. If there were no transients and
no other sources of systematic uncertainties, all the
curves would approach unity at large scales, where tree-level PT
applies. Unfortunately, there are other sources of systematic
uncertainties as we shall discuss below.

The different symbols correspond to different outputs of the
simulation: open triangles denote initial conditions ($a=1$, $\sigma_8
= 0.24$), solid triangles ($a=1.66$, $\sigma_8=0.40$), open squares
($a=2.75$, $\sigma_8=0.66$), and solid squares ($a=4.2$,
$\sigma_8=1.0$). The results for $S_3$ in the top left panel are those
presented in Fig.~9 of BGE. For the initial conditions measurements  
(open triangles) there is some disagreement with the ZA predictions,
especially at small scales, due to discreteness effects, which have not been
corrected for. The initial particle arrangement is a grid, therefore
the Poisson model commonly used to correct for discreteness is not
necessarily a good approximation (see BGE for further discussion of
this point). The second output time (solid triangles) is perhaps the
best for testing the 
predictions of transients: discreteness corrections become much smaller
due to evolution away from the initial conditions, and the system has not
yet evolved long enough so that finite volume corrections are 
important. For $S_3$ we see excellent agreement with the
predictions of Eq.~(\ref{S3}), with a small excess at small scales due
to non-linear evolution away from the tree-level prediction. For $p>3$
the numerical results show a similar behavior with 
increased deviation at small scales due to non-linear evolution, as
expected (see discussion below). For the last two outputs we see a
further increase of non-linear effects at small scales, then a
reasonable agreement (at least for $S_3$ and $S_4$) with the
transients predictions, and lastly a decrease of the numerical results
compared to the PT predictions at large scales due to finite volume
effects, which increase with $\sigma_8$, $R$, and $p$ (Colombi,
Bouchet \& Schaeffer 1994; BGE; Colombi, Bouchet \& Hernquist 1996,
Munshi et al. 1997). 
For reasons of clarity, only measurements with reasonable error bars
have been included, essentially as in BGE (see their Fig.~13).

The $N$-body results show a systematic overestimate of the PT 
predictions at small scales, i.e. $R \la 10$ h$^{-1}$Mpc for the last
two outputs. Here we should stress 
that the PT predictions in this paper correspond to tree-level
quantities, i.e. valid in the limit of vanishing variance, $\sigma^2(R)
\rightarrow 0$. For a small variance, one can use one-loop PT to
calculate the correction to the tree-level results, and neglecting
transients one finds that in general (for $n<-1$ in the scale-free case)

\beq
S_p(R)= S_p^{(0)} + \sigma^2(R) \ S_p^{(2)} + \ldots,
\label{Sp1L}
\eeq
where $S_p^{(0)}$ denotes the tree-level value and $S_p^{(2)}$ the
one-loop correction coefficient (Scoccimarro \& Frieman 1996). For
$p=3$ and top-hat smoothing, 
$S_3^{(0)}=3.86$, $S_3^{(2)}=3.18$ for $n=-2$ (Scoccimarro 1997). For
$p=4$ there are no available PT results to one-loop, but in the 
spherical collapse approximation (that has been recently extended to 
take into account loop-corrections by Fosalba \& Gazta\~naga (1997)), 
$S_4^{(0)}=27.56$, $S_4^{(2)}=63.56$ for $n_{\rm eff}= -2$.  This 
approximation neglects tidal effects, but comparison with numerical 
simulations and exact PT  results shows a  good 
agreement for the $S_{p}$ parameters (Fosalba \& Gazta\~naga 1997). 
For example, the skewness one-loop coefficient
$S_3^{(2)}=3.21$ for $n=-2$ is in excellent agreement with the exact PT 
result quoted above. These results show
that at scales of a few h$^{-1}$Mpc for CDM models, where $n_{\rm
eff} \approx -2$, one-loop corrections play a significant role even
for $\sigma^2(R) <1$, and increasingly with $p$ (Fosalba \&
Gazta\~naga 1997). This is the reason for the  excess of
scale dependence at $R \la 10$ h$^{-1}$Mpc in Fig.~\ref{fig5}.
Another interesting point of this
exercise with one-loop corrections is to show how a late start
in a simulation can mimic a ``false agreement'' with tree-level PT:
transients tend to decrease the measured $S_p$'s, whereas one-loop
corrections due to the finite variance tend to cancel this
decrease. This may lead to the illusion that tree-level PT has a wider
range of applicability than is really the case.

In Figure~\ref{fig5}, the PT calculations include the full dependence
on  $\gamma_p(R)$ parameters. An approximation  is sometimes used to 
calculate $S_{p}$ for $p>3$ in
which $\gamma_p(R)=0$ for $p \geq 2$, on the grounds that this is true
for scale-free spectra and CDM models have a slowly varying spectral
index. Figure~\ref{fig6} addresses the validity of this approximation
for standard CDM models. The top left panel 
shows $n_{\rm eff}=\gamma-3$ and the  $\gamma_p(R)$ parameters for $2
\leq p \leq 5$, whereas the top right panel shows the ratio of $S_p$ parameters
($4 \leq p \leq 7$) calculated using $\gamma_p(R)=0$ for $p \geq 2$ to
the full calculation. We see that the approximation works quite well
at small scales, but as $R$ increases (and the regime where
tree-level PT holds is reached) the approximation $\gamma_p(R)=0$ gradually
breaks down. The bottom panels in Fig.~\ref{fig6} show the same
calculation for a CDM model with shape parameter $\Gamma=0.21$ (which
just corresponds to a shift in scale with respect to the $\Gamma=0.5$
model), and the corresponding calculation in the standard CDM model
for the $T_p$ parameters. In this latter case, the approximation is
worse than for the $S_p$ parameters. Note that we have suppressed the
plotting of $T_p$
ratios beyond the point where they become negative for reasons of
clarity. 

The next two figures, Figs.~\ref{fig7}~and~\ref{fig8}, show the
predictions of transients as a function of scale factor $a$ for $S_p$
and $T_p$ parameters respectively, 
at smoothing scales $R=10,100$ h$^{-1}$Mpc (left and right panels,
respectively) for $\Gamma=0.5$ and $0.21$ CDM models (top and bottom panels, 
respectively). These calculations include the full dependence on
$\gamma_p(R)$ parameters, and correspond to ZA initial velocities. A
similar set of plots is presented in Figs.~\ref{fig9}~and~\ref{fig10}
for the case of velocities set as in EDFW. Comparing these two set of
plots we arrive to a similar 
conclusion than for scale-free spectra, velocities set from perturbed
particle positions lead to a faster rate of convergence to the
asymptotic exact values than just pure ZA initial velocities. The 
latter requires a factor of about two to three more 
expansion away from the initial conditions to reduce transients in
density and velocity statistics by the same amount. We also 
see that for $R=100$ h$^{-1}$Mpc, the 
transients are more important than for $R=10$ h$^{-1}$Mpc, as expected
from the results in the scale-free case. 
In the bottom panels, we include upper
horizontal axes which denote the scale factor $a$ for a cosmological
model in which $\Omega(a) =0.3$. This serves to illustrate the fact
that for models with $\Omega<1$, transients persist  longer because the
growth of fluctuations is governed by the linear growth factor which
evolves more slowly than the scale factor. Therefore, to achieve the same
accuracy regarding transient behavior, $\Omega<1$ models should be
started at a higher redshift than $\Omega=1$ simulations. The results
in Figs.~\ref{fig7}-\ref{fig10} show that an $\Omega=0.3$
model should be started at about a factor of two larger in redshift
than an $\Omega=1$ simulation. We remind the reader that this result is
approximate; it depends on the assumption that $f^2/\Omega \approx 1$,
but this approximation is better than 25\% for $\Omega \geq 0.3$.

Finally, in view of these results, one is led to ask whether it is
possible to decrease the magnitude of transients besides
the obvious solution of starting the simulation early enough. A
natural candidate to improve upon the ZA to set the initial conditions 
is to use second-order Lagrangian PT (hereafter 2LPT, e.g. Moutarde
et al. 1991, Buchert 1992, Bouchet et al. 1995). This procedure
requires minimal additional computational cost over the standard ZA
scheme, as discussed in detail in Appendix D. Since 2LPT reproduces 
growing modes of density and velocity perturbations to second-order, 
there are no transients
in the evolution of  $S_3$ and $T_3$ parameters in this
approximation. Figures~\ref{fig11} and~\ref{fig12} present the
predictions for transients from 2LPT initial conditions 
for different spectral indices as a function
of scale factor $a$ for $3 \leq p \leq 8$. The details of this
calculation and the analytic results are summarized in
Appendix~C. Compared to the corresponding results in Figs.~\ref{fig1} 
and~\ref{fig2} respectively (note the difference in scales), we see
that 2LPT initial conditions lead to an improvement of more than an
order of magnitude in the amount of expansion necessary to erase
transients over the standard ZA scheme. Moreover, it yields about a factor
of four improvement with respect to ZA initial conditions with
velocities set from perturbed particle positions; that is, a simulation
started with 2LPT initial conditions at redshift $z_{\rm start}=10$
would roughly correspond to a simulation started at $z_{\rm start}=40$
using the ZA with velocities as in EDFW, and $z_{\rm start}=100$ for a
standard ZA procedure. Given that generating EDFW velocities and 2LPT
initial conditions 
require similar additional computational costs over the 
ZA scheme (see Appendix~\ref{appD}), in any case very small compared 
to the actual cost of running the simulation,
2LPT seems the best alternative to standard ZA methods.

%
%
%===================================
\section{Conclusions and Discussion}
\label{conc}
%===================================
%
%

In this paper we give a perturbative analysis of  
the problem of transients from initial conditions when 
measuring moments of density and velocity fields in
numerical simulations, where initial conditions are 
usually set using the Zel'dovich approximation (ZA). Although the ZA
correctly reproduces the linear growing modes of density and velocity
perturbations, non-linear growth is inaccurately
represented by the ZA, because of the ZA failure to conserve momentum. 
This implies
that it takes a non-negligible amount of time for the correct dynamics
to establish the 
expected tree-level correlation hierarchy predicted by perturbation
theory at large scales. 

We focussed on one-point cumulants of the
density and velocity divergence fields, characterized by the
$S_p$ and $T_p$ parameters. We extended the standard formulation of
perturbation theory  to include to arbitrary
order the transient behavior of non-linear  
solutions  that encode the information on the amount of time
needed to overcome the correlations imprinted by the initial
conditions.  
Using these results, we calculated  the full
time-evolution of $S_{p}(a)$ and $T_{p}(a)$ to tree-level with
top-hat smoothing as a function of scale factor $a$. These results
interpolate between the initial values set by 
the ZA and the asymptotic values expected from the exact dynamics at
large scales. More importantly, they provide a quantitative estimation
of the magnitude of the effect and the amount of expansion needed to
achieve a given accuracy in the determination of moments of the
density and velocity fields. Needless to say, there
are many other uncertainties when measuring statistics such as
one-point cumulants in
numerical simulations which should be properly taken into account;
this has been extensively discussed in the literature
(e.g. Colombi, Bouchet \& Schaeffer 1994, 1995; BGE; Colombi, Bouchet
\& Hernquist 1996; Szapudi \& Colombi 1996; Munshi et al. 1997). 

We found that the magnitude of transients is determined, at a given
order $p$, by the effective spectral index $n_{\rm eff}$. If initial
conditions  are set at $a_{0}\equiv 1$, obtaining the skewness
factor $S_3$ ($T_3$) within 10$\%$ accuracy requires $a \approx 6$ ($a
\approx 15$) for $n_{\rm eff} \approx 0$, whereas higher (lower)
$n_{\rm eff}$ demands more (less) expansion away from initial conditions. 
Furthermore, these requirements become much more stringent as $p$
increases, always showing slower decay of transients for $T_p$ than
$S_p$, due to velocity correlations being poorly represented by the ZA. 
For models with density parameter $\Omega < 1$, the transients take
more expansion factors to die out, since the
relevant dynamical quantity that controls the growth of structure is
the linear growth factor, which evolves more slowly than the scale
factor. This implies, for example,  that an open model
where the final state corresponds to $\Omega=0.3$  requires roughly a
factor of two larger expansion away from the initial conditions to
erase transients than an $\Omega=1$ model. 
Thus, in general, numerical simulations of  models 
with $\Omega < 1$ should be started at higher redshift
than critical density models to reduce transients by the same amount. 
We have also explored the influence of
setting initial velocities on the magnitude of transients, and found
that velocities set as in EDFW from the gravitational potential due to
perturbed particle positions (i.e., after the ZA displacement has been
applied), the time-scale of transients is reduced by a factor of two
or three depending on the spectrum. 

The results of the predicted transients in $S_p$ for $3 \leq p \leq 6$ were 
compared with measurements in standard CDM numerical
simulations by Baugh, Gazta\~naga \& Efstathiou (1995). We found good
agreement, especially for intermediate output times where discreteness
and finite volume effects are not important and tree-level PT applies
over a wider range of scales. Our results show that the dependence of
transients on spectral index is opposite to that of finite volume
effects, which decrease with $n_{\rm eff}$ (e.g., Colombi, Bouchet \&
Hernquist 1996; Munshi et al. 1997).   
In the comparison shown in Fig.~\ref{fig5} we see that at
late times finite volume effects start to dominate over
transients, systematically decreasing $S_5(R)$ and $S_6(R)$ for $R \ga
20$ h$^{-1}$Mpc. However, as the size of
a CDM simulation box is increased and higher spectral indices are
probed, transients eventually dominate over finite volume effects,
since the latter become quite small as $n_{\rm eff} \rightarrow 1$. 

Another interesting issue regarding transients in numerical
simulations is  to investigate to what
extent transients are present at smaller scales. As $R$ decreases, the
PT approach used here breaks down (as shown in Fig.~\ref{fig5} for
$R \la 10$ h$^{-1}$Mpc in the last two outputs) since we use leading
order (tree-level) PT, which yields the $S_p$ and
$T_p$ parameters in the limit of vanishing variance. However, at least
in the intermediate regime where the variance is of order unity and
one-loop PT works well (Scoccimarro 
\& Frieman 1996; Scoccimarro 1997), the corrections to the
predictions in this paper take the
form given by Eq.~(\ref{Sp1L}). In this case, the expected result is
that transients {\em will take longer to decay} than for tree-level
quantities because loop-corrections depend on higher-order
non-linearities which are increasingly underestimated by the ZA. That is in
fact the reason why transients in $S_p$ and $T_p$ take longer to decay
as $p$ increases. On the other hand, for CDM models the spectral
index decreases at smaller scales, and the magnitude of transients will
decrease. As smaller scales are approached and the variance
becomes larger than unity, shell-crossing becomes important and it is
less clear what to expect, this depends on how much the dynamics of
virializing high-density regions couples to the large-scale modes.

Since the results of transients in this work concern the
statistical properties of density and velocity fields, it is likely
that other statistical measures of clustering will show a similar
effect, in particular those sensitive to phase correlations,
where non-linear dynamics provides the leading contribution. An obvious
candidate closely related to the discussion in this paper is the
measurement of one-point cumulants in redshift space (Hivon et al. 1995), 
which can be thought as appropriate cross-correlations between density
and velocity fields (Scoccimarro, Couchman \& Frieman 1997). Based on
the present results, one  expects in this case that the $S_p$ parameters in
redshift space will be more affected by transients that the
corresponding ones in real space, since velocity correlations are more
affected by transients. Other statistics
sensitive to large-scale phase coherence, such as percolation studies,
may show similar effects regarding transients to the ones investigated
in this work. 

The effect of transients also implies that one must be very
careful when comparing non-linear approximations and perturbative
results to numerical simulations which have not been evolved for a
long enough time. In the former case, comparison among non-linear
approximations usually concludes that the ZA (or modifications thereof)
is a good approximation to the fully non-linear numerical
simulation. These conclusions should be taken with caution if the 
simulation has not been started early enough so that 
transients from the initial conditions are still present. Also, as
discussed in Section~\ref{res}, when comparing
numerical simulations to PT predictions, the effect of transients
combined with finite values of the variance characteristic of a late
simulation start 
tends to create the false impression that tree-level PT has a wider
range of validity than is really the case.

The results in this paper should perhaps be viewed as a useful
guideline for designing numerical simulations with interest in
measuring higher-order statistics of density and velocity fields, as
well as other measures of clustering.  The main conclusion 
in this regard is the requirement for an early enough 
start to avoid undesired effects from transients, 
particularly for studies of clustering statistics 
at high redshift. Although an early start does
not cause significant overhead in the time required to run a
simulation because early time-steps run quite rapidly due to weak
clustering, there are reasons to avoid starting a simulation at very
high redshift. The faster Hubble expansion rate at earlier times demands
successively shorter time-steps, but more importantly, the numerical
integration of the equations of motion leads to suppressed growth 
of small-scale modes that becomes more significant by starting at higher
redshifts (Couchman 1997). For these reasons, it is desirable to use a
better approximation to  generate initial conditions. 
A natural candidate is second-order Lagrangian
PT (2LPT), which reproduces growing modes to second-order in
non-linear PT. As shown in this work, this can reduce
the expansion required to decay 
away transients by more than one order of magnitude compared to the
standard ZA scheme. This is a significant improvement, particularly
given that its implementation is simple and only requires minimal inexpensive
modifications to widely used ZA initial conditions codes.

% What is the effect on statistics measured in non-linear regime? Once
% things shell cross, expect negligible contribution from initial
% conditions. However, for voids things can be quite different, as it's
% usually noted for grid effects.
% Wait a minute.. The concept of propagator extends to non-linear 
% behavior too. Expect similar magnitude of transients. RG actually 
% is a nice way of getting rid of transients and allowing dynamics 
% to relax into its final state.

% Comment on simulations in the literature that studied $S_p$ and $T_p$,
% when did they start? Bernardeau \& Kofman test of PDF? Higher $S_p$'s
% are more affected by transients, expect PDF to be very sensitive to
% this then. LOOK AT EDGEWORTH EXPANSION => TRANSIENTS FOR PDF
% Conspiracy? Lokas et al., only people to start early
% enough? 

% Easy way to predict time-scale as a function of $p$?

% What about non-Gaussian initial conditions? Do they take care of
% transients properly? Results in this paper relevant for any models
% with quasi-hierarchical non-Gaussian initial conditions.

% Review of PT, agreement with simulations, interplay PT-$N$-body, etc. 

% Importance of $S_p$ and $T_p$ as statistics: non-Gaussianity, $\Omega$
% dependence, bias. Lensing? $T_p$ proposed as a measure of $\Omega$
% (references).

% Munshi \& Starobinsky, and others comparing non-linear approx to ED.

\acknowledgements   
This work was in part motivated by a conversation with Hugh Couchman
and P.J.E.  Peebles. I would also like to  thank St\'ephane
Colombi, Pablo Fosalba, Josh Frieman,  
Dmitry Pogosyan, and Istv\'an Szapudi for comments and discussions, and Enrique 
Gazta\~naga  for providing numerical  
simulation measurements and for comments as well. Additional thanks
are due to Hugh Couchman for numerous helpful discussions . I thank the Aspen
Center for Physics for hospitality during the workshop 
``Precision Measures of Large-Scale Structure'', where this project
was started. 

\clearpage

\appendix
%
%
%===============================================================
\section{Tree-Level Recursion Relations for Smoothed Vertices including Transients}
\label{appA}
%===============================================================
%
%

We define the angular averaged smoothed kernels $\omega_{a}^{(n)}$:

\beq
\omega_{a}^{(n)} \equiv \int \ \Big(\frac{d\Omega_{1}}{4\pi}\Big) \ldots 
\Big(\frac{d\Omega_{n}}{4\pi}\Big) \ W_{1\ldots n} \ {\cal 
F}_{a}^{(n)}(\k_{1},\ldots,\k_{n}),
\eeq
and similarly for the angular averaged  
initial conditions kernels $\eta_{a}^{(n)}$ upon 
replacing ${\cal F}_{a}^{(n)}$ by ${\cal I}_{a}^{(n)}$. 
As shown in detail by Bernardeau (1994), the operation of smoothing 
one-point cumulants with top-hat
windows can be easily calculated due to special geometric
properties. In terms of the scattering matrix elements in
Eq.~(\ref{scatt-m}), the following properties of angular
integration hold 

\bml
\label{THsmoothing}
\beqa
\int \Big(\frac{d \Omega_1}{4\pi}\Big)\  \Big(\frac{d
\Omega_1}{4\pi}\Big)  \ \gamma_{121}(\k,\k_1,\k_2) \ W_{12} &=& 
W_1 \ W_2 + \frac{1}{3} \ W_1 \ k_2 R \ W_2' \\
\nonumber \\ 
\int \Big(\frac{d \Omega_1}{4\pi}\Big)\  \Big(\frac{d
\Omega_1}{4\pi}\Big)  \ \gamma_{222}(\k,\k_1,\k_2) \ W_{12} &=&
\frac{1}{3} \ W_1 \ W_2 + \frac{1}{6} \ W_1 \ k_2 R \ W_2' + \frac{1}{6}
\ W_2 \ k_1 R \ W_1' , 
\eeqa
\eml
where the prime denotes a derivative (Bernardeau 1994). To prove these
results, all we need is the expansion of top-hat windows such as $W_{12}$, in
terms of $W_1$, $W_2$ (and their derivatives), and Legendre polynomials
of the angle between $\k_1$ and $\k_2$. This, it turns out, is a
straightforward application of Gegenbauer's addition theorem for
Bessel functions (e.g., Watson 1944).   Then, using Eq.~(\ref{THsmoothing})
and the recursion relations in Eq.~(\ref{pt_recrel_sd}) we find:

\beqa
\label{verticesRR}
\omega_{a}^{(n)}(z) &=& 
{\rm e}^{-n z} \ g_{ab}(z) \ \eta_{b}^{(n)} + 
\sum_{m=1}^{n-1}
\int_0^z ds  \ {\rm e}^{n(s-z)} \Bigg\{ \frac{1}{3} \ g_{a1}(z-s) \ 
\Big[ 3 \ \omega_2^{(m)} \ \omega_1^{(n-m)} + 
\omega_2^{(m)} \ \partial_{\kappa}\omega_1^{(n-m)}  \Big]  
\nonumber \\ & & + \frac{1}{6} \ g_{a2}(z-s) \ 
\Big[2 \  \omega_2^{(m)} \ \omega_2^{(n-m)} + 
\omega_2^{(m)} \ \partial_{\kappa }\omega_2^{(n-m)}  + 
 \partial_{\kappa}\omega_2^{(m)} \ \omega_2^{(n-m)}  \Big] 
\Bigg\},
\eeqa
where $\kappa \equiv \ln R$ and the quantities in square brackets are 
evaluated at time $s$. This recursion relation (and its derivatives) 
can be used to obtain the smoothed vertices to arbitrary order in PT 
in terms of top-hat window functions and their derivatives with 
respect to scale, which are then straightforwardly converted to 
derivatives of the variance with respect to scale, yielding the $\gamma_p(R)$
parameters defined in Eq.~(\ref{gamma_p}).
If we ignore transients, and assume Gaussian initial conditions,
Eq.~(\ref{verticesRR}) reduces to ($n>1$)

\beqa
\label{verticesRRgNT}
\omega_{a}^{(n)}  &=&  
\sum_{m=1}^{n-1}
 \Bigg\{ \frac{1}{3} \  \sigma_{a1}(n)  \ 
\Big[ 3 \ \omega_2^{(m)} \ \omega_1^{(n-m)} + 
\omega_2^{(m)} \ \partial_{\kappa}\omega_1^{(n-m)}  \Big]  
\nonumber \\ & & + \frac{1}{6} \  \sigma_{a2}(n)  \ 
\Big[2 \  \omega_2^{(m)} \ \omega_2^{(n-m)} + 
\omega_2^{(m)} \ \partial_{\kappa }\omega_2^{(n-m)}  + 
 \partial_{\kappa}\omega_2^{(m)} \ \omega_2^{(n-m)}  \Big] 
\Bigg\},
\eeqa
with 

\beq
\sigma_{ab}(n)  \equiv \frac{1}{(2n+3)(n-1)}
\Bigg[ \begin{array}{cc} 2n+1 & 2 \\ 3 & 2n \end{array} \Bigg].
\label{sigma}
\eeq
If we further assume no smoothing, Eq.~(\ref{verticesRRgNT}) yields

\beq 
\label{verticesRRgNTus}
\omega_{a}^{(n)}  =   \sigma_{ab}(n)\ \bar\gamma_{bcd} 
\sum_{m=1}^{n-1} \omega_c^{(m)} \ \omega_d^{(n-m)},
\eeq 
where $\bar\gamma_{121} =1$, $\bar\gamma_{222} =1/3$, 
and zero otherwise, which is the limit of Eq.~(\ref{THsmoothing}) as
the smoothing scale goes to zero, $R \rightarrow 0$. This simple
recursion relation is an alternative method to reproduce the
well-known results for unsmoothed vertices usually obtained using the
spherical collapse approximation, i.e. $\nu_2=34/21$, $\nu_3=682/189$,
$\nu_4=446440/43659$, \ldots, 
$\mu_2=26/21$, $\mu_3=142/63$, $\mu_4=236872/43659$, and so on
(Bernardeau 1992).

% Formulas used, gegenbauer, so on. 2D? Briefly comment on {\sf
% Mathematica} code. Physical interpretation of geometrical properties
% of top-hat windows.

%
%
%===============================================================
\section{Results for Transients in Kurtosis and Pentosis Factors}
\label{appB}
%===============================================================
%
%
In this appendix, we present results for the transients in $S_4, T_4,
S_5$ and $T_5$ from ZA initial conditions, for standard ZA velocities
and initial velocities as in EDFW, set from
perturbed particle positions (Eq.~(\ref{vel-corr2})). 
For $p=4$, evaluation of Eq.~(\ref{S4res}) using the methods described
in Appendix A leads to  

\beqa
\label{p=4za}
S_4(a) &=& {{60712}\over {1323}} - {{62\ \gamma}\over 3} 
+ {{7\ {\gamma^2}}\over 3} - {{2\ {\gamma_2}}\over 3} 
- {{816}\over {35\,a}} + {{28\ \gamma}\over {5\,a}}+ {{184}\over {75\,{a^2}}} 
+ {{1312}\over {245\,{a^{{7\over 2}}}}} - {{8\ \gamma}\over {5\,{a^{{7\over2}}}}} 
-{{1504}\over {4725\,{a^{{9\over2}}}}}+ {{192}\over {1225\,{a^7}}} ,\nonumber \\
& &  \\ 
T_4(a) &=& 
{{12088}\over {441}} - {{338\ \gamma}\over {21}} +   {{7\ {\gamma^2}}\over 3} 
- {{2\ {\gamma_2}}\over 3}-   {{624}\over {35\,a}} + {{28\ \gamma}\over {5\,a}} 
+ {{184}\over {75\,{a^2}}} - {{1192}\over {245\,{a^{{7\over2}}}}} 
+   {{64\ \gamma}\over {35\,{a^{{7\over2}}}}} 
+   {{752}\over {1575\,{a^{{9\over 2}}}}} + {{432}\over {1225\,{a^7}}},\nonumber \\
& &
\eeqa
where the $\gamma_p$ parameters are defined in Eq.~(\ref{gamma_p}).
For $a=1$, these expressions reduce to the initial kurtosis factors
given by the ZA, and at late times they approach the asymptotic exact
values (Bernardeau 1994). Similarly, for EDFW initial velocities we
obtain instead 

\beqa
\label{p=4}
S_4(a) &=& S_4(\infty)
-   {{272}\over {35\,a}} + {{28\ \gamma}\over {15\,a}} 
+ {8\over {225\,{a^2}}} -   {{5248}\over {735\,{a^{{7\over 2}}}}}
+ {{32\ \gamma}\over {15\,{a^{{7\over 2}}}}}
- {{5056}\over {4725\,{a^{{9\over 2}}}}} + {{1024}\over {3675\,{a^7}}},\\ 
T_4(a) &=& T_4(\infty)-   {{208}\over {35\,a}} + {{28\ \gamma}\over {15\,a}} 
+ {8\over {225\,{a^2}}} 
+   {{4768}\over {735\,{a^{{7\over 2}}}}} 
- {{256\ \gamma}\over {105\,{a^{{7\over 2}}}}} + {{2528}\over
{1575\,{a^{{9\over 2}}}}}  + {{768}\over {1225\,{a^7}}},  
\eeqa
where $S_4(\infty)$ and $T_4(\infty)$ denote the asymptotic
exact-dynamics values given by the first four terms in
Eqs.~(\ref{p=4za}). 
For $p=5$, Eq.~(\ref{S5res}) yields for the pentosis parameters from
standard ZA initial velocities

\beqa
S_5(a) &=& {{200575880}\over {305613}}- {{1847200\ \gamma}\over {3969}} 
+ {{6940\ {\gamma^2}}\over {63}} - {{235\ {\gamma^3}}\over {27}} 
- {{1490\ {\gamma_2}}\over {63}} + {{50\ \gamma\ {\gamma_2}}\over 9} 
- {{10\ {\gamma_3}}\over {27}}
\nonumber \\&- &  {{74584}\over {147\,a}} 
+ {{252\ \gamma}\over a}- {{94\ {\gamma^2}}\over {3\,a}} 
+ {{20\ {\gamma_2}}\over {3\,a}} 
+ {{7208}\over {63\,{a^2}}} - {{1352\ \gamma}\over {45\,{a^2}}} 
- {{1384}\over {225\,{a^3}}}+ {{317104}\over {3087\,{a^{{7\over 2}}}}} 
- {{8984\ \gamma}\over {147\,{a^{{7\over 2}}}}} 
+ {{188\ {\gamma^2}}\over {21\,{a^{{7\over 2}}}}} 
\nonumber \\&- &  {{40\ {\gamma_2}}\over {21\,{a^{{7\over 2}}}}} 
- {{98656}\over {3969\,{a^{{9\over 2}}}}} 
+ {{3296\ \gamma}\over {405\,{a^{{9\over 2}}}}} 
+ {{17504}\over {17325\,{a^{{{11}\over 2}}}}} + {{1944}\over {343\,{a^7}}} 
- {{512\ \gamma}\over {245\,{a^7}}} + {{4352}\over {11025\,{a^8}}} 
+ {{768}\over {8575\,{a^{{{21}\over 2}}}}},\label{p=5za} \\ 
& & \nonumber \\ 
T_5(a) &=& 
{{94262120}\over {305613}}- {{161440\ \gamma}\over {567}} 
+ {{260\ {\gamma^2}}\over 3}- {{235\ {\gamma^3}}\over {27}} 
- {{130\ {\gamma_2}}\over 7}+ {{50\ \gamma\ {\gamma_2}}\over 9} 
- {{10\ {\gamma_3}}\over {27}} 
\nonumber \\&- & {{14792}\over {49\,a}} + {{4108\ \gamma}\over {21\,a}} 
- {{94\ {\gamma^2}}\over {3\,a}} + {{20\ {\gamma_2}}\over {3\,a}} 
+ {{5512}\over {63\,{a^2}}}- {{1352\ \gamma}\over {45\,{a^2}}} 
- {{1384}\over {225\,{a^3}}}- {{68888}\over {1029\,{a^{{7\over 2}}}}} 
+ {{1012\ \gamma}\over {21\,{a^{{7\over 2}}}}} 
- {{26\ {\gamma^2}}\over {3\,{a^{{7\over 2}}}}} 
\nonumber \\&+ & {{40\ {\gamma_2}}\over {21\,{a^{{7\over 2}}}}} 
+ {{92072}\over {3969\,{a^{{9\over 2}}}}}
- {{26248\ \gamma}\over {2835\,{a^{{9\over 2}}}}} 
- {{8752}\over {5775\,{a^{{{11}\over 2}}}}} 
+ {{1944}\over {343\,{a^7}}} 
- {{96\ \gamma}\over {35\,{a^7}}} 
+ {{1088}\over {1225\,{a^8}}} 
- {{2592}\over {8575\,{a^{{{21}\over 2}}}}}, \label{p=5za2}
\eeqa 
whereas for EDFW initial velocities we get

\beqa
\label{p=5}
S_5(a) &=& S_5(\infty) - {{74584}\over {441\,a}} + {{84\ \gamma}\over a} 
- {{94\ {\gamma^2}}\over {9\,a}}+ {{20\ {\gamma_2}}\over {9\,a}} 
+ {{136}\over {27\,{a^2}}}- {{184\ \gamma}\over {135\,{a^2}}} 
- {{352}\over {225\,{a^3}}}- {{1268416}\over {9261\,{a^{{7\over 2}}}}} 
+ {{35936\ \gamma}\over {441\,{a^{{7\over 2}}}}}
\nonumber \\&- &  {{752\ {\gamma^2}}\over {63\,{a^{{7\over 2}}}}} 
+ {{160\ {\gamma_2}}\over {63\,{a^{{7\over 2}}}}} 
- {{88960}\over {3969\,{a^{{9\over 2}}}}} 
+ {{2624\ \gamma}\over {405\,{a^{{9\over 2}}}}} 
- {{14272}\over {7425\,{a^{{{11}\over 2}}}}} 
+ {{3456}\over {343\,{a^7}}} 
- {{8192\ \gamma}\over {2205\,{a^7}}} 
+ {{99328}\over {33075\,{a^8}}} 
- {{16384}\over {77175\,{a^{{{21}\over 2}}}}}, \nonumber \\
& & \\
T_5(a) &=& T_5(\infty) - {{14792}\over {147\,a}} + {{4108\ \gamma}\over {63\,a}} 
- {{94\ {\gamma^2}}\over {9\,a}} + {{20\ {\gamma_2}}\over {9\,a}} 
+ {{104}\over {27\,{a^2}}}- {{184\ \gamma}\over {135\,{a^2}}} 
- {{352}\over {225\,{a^3}}}+ {{275552}\over {3087\,{a^{{7\over 2}}}}} 
- {{4048\ \gamma}\over {63\,{a^{{7\over 2}}}}} 
\nonumber \\&+ &  {{104\ {\gamma^2}}\over {9\,{a^{{7\over 2}}}}} 
- {{160\ {\gamma_2}}\over {63\,{a^{{7\over 2}}}}} 
+ {{89024}\over {3969\,{a^{{9\over 2}}}}} 
- {{24112\ \gamma}\over {2835\,{a^{{9\over 2}}}}} 
+ {{7136}\over {2475\,{a^{{{11}\over 2}}}}} 
+ {{3456}\over {343\,{a^7}}} - {{512\ \gamma}\over {105\,{a^7}}} 
+ {{24832}\over {3675\,{a^8}}} 
+ {{6144}\over {8575\,{a^{{{21}\over 2}}}}}, \nonumber \\
& &
\eeqa 
where $S_5(\infty)$ and $T_5(\infty)$ denote denote the asymptotic
exact-dynamics values given by the first line in
Eqs.~(\ref{p=5za}) and (\ref{p=5za2}). 

%
%
%===============================================================
\section{Results for Transients from Second-Order Lagrangian PT Initial Conditions }
\label{appC}
%===============================================================
%
%

To calculate the properties of initial conditions in 2LPT, it is
convenient to take advantage of the results by Munshi, Sahni \&
Starobinsky (1994), who derived the density and velocity divergence
vertex generating functions in 2LPT 
for the case in which smoothing effects are neglected. The effects of
smoothing can then be included by the mapping given by Bernardeau
(1994b). We refer the reader to these papers for details, here we just
present the results of these calculations including the effects of
transients. For simplicity, we just display results assuming
$\gamma_p=0$ for $p \geq 2$. The skewness parameters, $S_3$ and $T_3$
do not show any transient behavior in 2LPT, since these are reproduced
exactly. For the kurtosis factors we have:
 
\beq
S_4(a) = S_4(\infty) - {{184}\over {105\,{a^2}}}
+ {{736}\over {945\,{a^{{9\over 2}}}}}, 
\eeq
\beq
T_4(a) = T_4(\infty) - {{184}\over {105\,{a^2}}} 
-{{368}\over {315\,{a^{{9\over 2}}}}},
\eeq
whereas for $p=5$ we obtain:

\beqa
S_5(a) &= & S_5(\infty)- {{25024}\over {441\,{a^2}}}
+ {{920\ \gamma}\over {63\,{a^2}}}  
+   {{320}\over {441\,{a^3}}} 
+   {{85376}\over {3969\,{a^{{9\over 2}}}}} 
- {{3680\ \gamma}\over {567\,{a^{{9\over 2}}}}}  ,
+   {{12800}\over {4851\,{a^{{{11}\over 2}}}}} 
\eeqa
\beqa
T_5(a) &= & T_5(\infty)- {{19136}\over {441\,{a^2}}}
+ {{920\ \gamma}\over {63\,{a^2}}}+ {{320}\over {441\,{a^3}}}
- {{83536}\over {3969\,{a^{{9\over 2}}}}}   
+ {{4600\ \gamma}\over {567\,{a^{{9\over 2}}}}} 
- {{6400}\over {1617\,{a^{{{11}\over 2}}}}}  .
\eeqa
Finally, the case $p=6$ yields:

\beqa
S_6(a) &=& S_6(\infty)- {{16275904}\over {9261\,{a^2}}} 
+ {{134872\ \gamma}\over {147\,{a^2}}} - {{7544\ {\gamma^2}}\over {63\,{a^2}}}
+ {{108800}\over {3087\,{a^3}}} - {{4160\ \gamma}\over {441\,{a^3}}}
+ {{50888}\over {2205\,{a^4}}}+ {{5567104}\over {9261\,{a^{{9\over2}}}}} 
\nonumber \\ & & 
- {{203872\ \gamma}\over {567\,{a^{{9\over 2}}}}}+ 
  {{30176\ {\gamma^2}}\over {567\,{a^{{9\over 2}}}}}
+ {{1280000}\over {11319\,{a^{{{11}\over 2}}}}} - 
 {{166400\ \gamma}\over {4851\,{a^{{{11}\over 2}}}}}  
- {{3786688}\over {601965\,{a^{{{13}\over 2}}}}}
+ {{67712}\over {19845\,{a^9}}},
\eeqa
\beqa
T_6(a) &=& T_6(\infty) - {{357696}\over {343\,{a^2}}}
+ {{313352\ \gamma}\over {441\,{a^2}}} - {{7544\ {\gamma^2}}\over {63\,{a^2}}}
+ {{83200}\over {3087\,{a^3}}} - {{4160\ \gamma}\over {441\,{a^3}}} 
+ {{50888}\over {2205\,{a^4}}}- {{11778944}\over {27783\,{a^{{9\over 2}}}}}
\nonumber \\ & & 
+ {{418784\ \gamma}\over {1323\,{a^{{9\over 2}}}}}
- {{33304\ {\gamma^2}}\over {567\,{a^{{9\over 2}}}}}
- {{3904000}\over {33957\,{a^{{{11}\over 2}}}}}
+ {{217600\ \gamma}\over {4851\,{a^{{{11}\over 2}}}}}
+ {{1893344}\over {200655\,{a^{{{13}\over 2}}}}}
+ {{16928}\over {2205\,{a^9}}}.
\eeqa

%
%
%===============================================================
\section{Lagrangian Perturbation Theory}
\label{appD}
%===============================================================
%
%

In this appendix we present the results needed to implement 2LPT
initial conditions in numerical simulations. These perturbative
results are not new 
and have been reported before in the literature (e.g. Buchert et al. 1994,
Bouchet et al. 1995), but they are collected and summarized here with 
emphasis on the practical issues to ease the implementation by interested
readers. In particular, Buchert et al. (1994) present the necessary
results to implement third-order Lagrangian PT (3LPT) solutions. 
However, a similar calculation to that in Appendix~\ref{appC} shows
that this only improves over 2LPT by a factor of two or so in the
expansion required to erase transients. Given the additional
complexity of 3LPT, which involves solving three additional Poisson
equations, it does not appear worthwhile to consider.

%
%===============================================================
\subsection{Basic Results of Second-Order Lagrangian PT}
\label{appD1}
%===============================================================
%

In Lagrangian PT, the object of interest is the displacement field
${\bf \Psi}(\q)$ which maps the initial particle positions $\q$ into the
final Eulerian particle positions $\x$, 

\beq
\x = \q + {\bf \Psi}(\q).
\eeq
The equation of motion for particle trajectories $\x(\tau)$ is

\beq
\frac{d^2 \x}{d \tau^2} + {\cal H}(\tau) \ \frac{d \x}{d \tau}= -
\del \Phi, 
\eeq
where $\Phi$ denotes the gravitational potential, and $\del$ the
gradient operator in Eulerian coordinates $\x$. Taking the
divergence of this equation we obtain a closed equation for the
displacement field

\beq
J(\q,\tau)\ \del \cdot \Big[ \frac{d^2 \x}{d \tau^2} + {\cal
H}(\tau) \ \frac{d \x}{d \tau} \Big] = \frac{3}{2} \Omega {\cal H}^2
(J-1),
\label{leom}
\eeq
where we have used Poisson equation together with the fact that
$1+\d(\x) =J^{-1}$, and 
the Jacobian $J(\q,\tau)$ is the determinant 

\beq
J(\q,\tau)  \equiv  {\rm Det}\Big( \d_{ij}+ \Psi_{i,j} \Big),
\eeq
where $\Psi_{i,j} \equiv \partial
\Psi_i /\partial \q_j$. Equation~(\ref{leom}) can be fully rewritten
in terms of 
Lagrangian coordinates by using that $\del_i = ( \d_{ij}+
\Psi_{i,j})^{-1} \del_{q_j}$, where $\del_q \equiv \partial /\partial \q$
denotes the gradient 
operator in Lagrangian coordinates. The resulting non-linear equation for
${\bf \Psi}(\q)$ is then solved perturbatively, expanding about its linear
solution, the Zel'dovich (1970) approximation 

\beq
\del_q \cdot {\bf \Psi}^{(1)}= -D_1(\tau) \ \d(\q),
\label{Psi1}
\eeq
which incorporates the kinematic aspect of the collapse of
fluid elements in Lagrangian space. Here $\d(\q)$ denotes the
(Gaussian) density field imposed by 
the initial conditions and $D_1(\tau)$ is the linear growth factor,
which obeys Eq.~(\ref{D1}). The solution to second order describes the
correction to the ZA displacement due to gravitational tidal effects and 
reads  

\beq
\del_q \cdot {\bf \Psi}^{(2)}= \frac{1}{2} D_2(\tau) \sum_{i \neq j}
(\Psi_{i,i}^{(1)} \Psi_{j,j}^{(1)} - \Psi_{i,j}^{(1)} \Psi_{j,i}^{(1)}),
\label{Psi2}
\eeq
(e.g., Bouchet et al. 1995) where $D_2(\tau)$ denotes the second-order
growth factor, which for  $0.1 \leq \Omega \leq 3$ ($\Lambda=0$) obeys 

\beq
D_2(\tau) \approx -\frac{3}{7} D_1^2(\tau) \ \Omega^{-2/63} \approx
-\frac{3}{7} D_1^2(\tau) 
\eeq
to better than 0.5\% and 7\%, respectively (Bouchet et
al. 1995), whereas for flat models with non-zero cosmological
constant $\Lambda$ we have for  $0.01 \leq \Omega \leq 1$

\beq
D_2(\tau) \approx -\frac{3}{7} D_1^2(\tau) \ \Omega^{-1/143} \approx
-\frac{3}{7} D_1^2(\tau),  
\eeq
to better than 0.6\% and 2.6\%, respectively (Bouchet et al. 1995). 
Since Lagrangian solutions up to second-order are
curl-free, it is convenient to define Lagrangian potentials
$\phi^{(1)}$ and $\phi^{(2)}$ 
so that in 2LPT

\beq
\x(\q) = \q -D_1\ \del_q \phi^{(1)} + D_2\ \del_q \phi^{(2)},
\label{dis2}
\eeq
and the velocity field then reads ($t$ denotes cosmic time)

\beq
{\bf v} \equiv \frac{d \x}{d t} = -D_1\ f_1\ H\ \del_q \phi^{(1)}
+ D_2\ f_2\ H\ \del_q \phi^{(2)},
\label{vel2}
\eeq 
where $H$ is the Hubble constant, 
and the logarithmic derivatives of the growth factors $f_i \equiv 
(d\ln D_i)/(d \ln a)$ can be approximated for open models with
$0.1 \leq \Omega \leq 1$ by 

\beq
f_1 \approx \Omega^{3/5}, \ \ \ \ \ f_2 \approx 2 \ \Omega^{4/7},
\eeq 
to better than 2\% (Peebles 1976) and 5\% (Bouchet et al. 1995),
respectively. For flat models with non-zero cosmological
constant $\Lambda$ we have for  $0.01 \leq \Omega \leq 1$

\beq
f_1 \approx \Omega^{5/9}, \ \ \ \ \ f_2 \approx 2 \ \Omega^{6/11},
\eeq 
to better than 10\% and 12\%, respectively (Bouchet et al. 1995). The
accuracy of these two fits improves significantly for $\Omega \geq
0.1$, in the range relevant for the present purposes.  
The time-independent potentials in Eqs.~(\ref{dis2}) and~(\ref{vel2}) obey
the following Poisson equations (Buchert et al. 1994)

\bml
\label{poisson2lpt}
\beqa
\del_q^2 \phi^{(1)}(\q) &=&  \d(\q), 
\label{phi1} \\ 
\del_q^2 \phi^{(2)}(\q)&=&  \sum_{i>j}
[\phi_{,ii}^{(1)}(\q)\ \phi_{,jj}^{(1)}(\q) - (\phi_{,ij}^{(1)}(\q))^2],
\label{phi2}
\eeqa
\eml

%
%===============================================================
\subsection{Setting up Second-Order Lagrangian PT Initial Conditions }
\label{appD2}
%===============================================================
%

{\em We now have all the elements to describe a simple prescription to
implement 2LPT initial conditions} (hereafter, a tilde denotes a Fourier-space
quantity): 

- A Gaussian density field, $\tilde\d(\k)$ is generated in Fourier space
from the desired linear power spectrum, and therefore
$\tilde\phi^{(1)}(\k)$ follows from Eq.~(\ref{phi1}) in Fourier space.

- Inverse fast Fourier transform (FFT) of $\tilde\phi^{(1)}(\k)$ yields
$\phi^{(1)}(\q)$ on the grid of unperturbed particle positions. 

- $\del_q \phi^{(1)}$ is obtained by differencing
$\phi^{(1)}(\q)$ along the three directions, giving the necessary
ingredients to displace 
particles and assign velocities according to the ZA. So far,
this is the usual procedure.

- The array $\del_q \phi^{(1)}$ can be stored temporarily in the array reserved
for the velocities assignment. Then we need three additional arrays of
dimension $N_{\rm grid}$ (where $N_{\rm grid}$ is the dimension of the
grid used to set up the initial conditions, usually equal to the number
of particles) to calculate the source term in Eq.~(\ref{phi2}). By
differencing the components of the array $\del_q \phi^{(1)}$ in a
diagonal fashion we obtain
the diagonal terms $\del^2_{11}\phi^{(1)}$, $\del^2_{22}\phi^{(1)}$,
and $\del^2_{33}\phi^{(1)}$, each stored in a $N_{\rm grid}$ array,
which are then multiplied together
to give the first contribution to the source term in
Eq.~(\ref{phi2}). The second contribution, consisting of the
non-diagonal terms $\phi_{,ij}^{(1)}(\q)$ in Eq.~(\ref{phi2}), is
obtained by simply 
differencing and accumulating in turn, without the need of additional
memory space. This yields the source term to the second Poisson
equation on the grid of unperturbed particle positions, stored in an
array of dimension $N_{\rm grid}$.

- FFT of this source term, solving Eq.~(\ref{phi2}) for
$\tilde\phi^{(2)}(\k)$ in Fourier
space, and inverse FFT yields the second-order potential
$\phi^{(2)}(\q)$ on the grid of unperturbed particle positions.

- The $\phi^{(2)}(\q)$ array is then differenced along one  
direction (recycling one of the two usable $N_{\rm grid}$ arrays to
thus store $\del_1\phi^{(2)}$) and then particles are displaced and
velocities assigned along the direction in question by combining
$\del_1\phi^{(1)}$ and $\del_1\phi^{(2)}$ according to
Eqs.~(\ref{dis2}) and~(\ref{vel2}). The same 
procedure is applied in turn to the two remaining directions.

We therefore see that the additional requirements for 2LPT initial
conditions generation over the standard ZA scheme is basically two
FFT's, a $3 \times N_{\rm grid}$ array, and differencing of the
Lagrangian potentials. These are indeed very modest by speed and
memory considerations, especially given the fact that setting up 
initial conditions is always a small fraction of the cost of running
the numerical simulation. In view of the substantial improvement (more
than one order of magnitude) on the required amount of expansion to
erase transients from initial conditions, it seems well worth the
minimal extra effort.

\clearpage

%
%
%=================

\clearpage 
%%%%%%%%%%%%%%%%%%%%%%%%%%%%%%%%%%%%%%%%%%%%%%%%%%%%%%%%%%%%%%%%%%%%%%%%%%%%%
%			FIGURES
%%%%%%%%%%%%%%%%%%%%%%%%%%%%%%%%%%%%%%%%%%%%%%%%%%%%%%%%%%%%%%%%%%%%%%%%%%%%%

\begin{figure}[t!]
\centering
\centerline{\epsfxsize=18truecm\epsfysize=18truecm\epsfbox{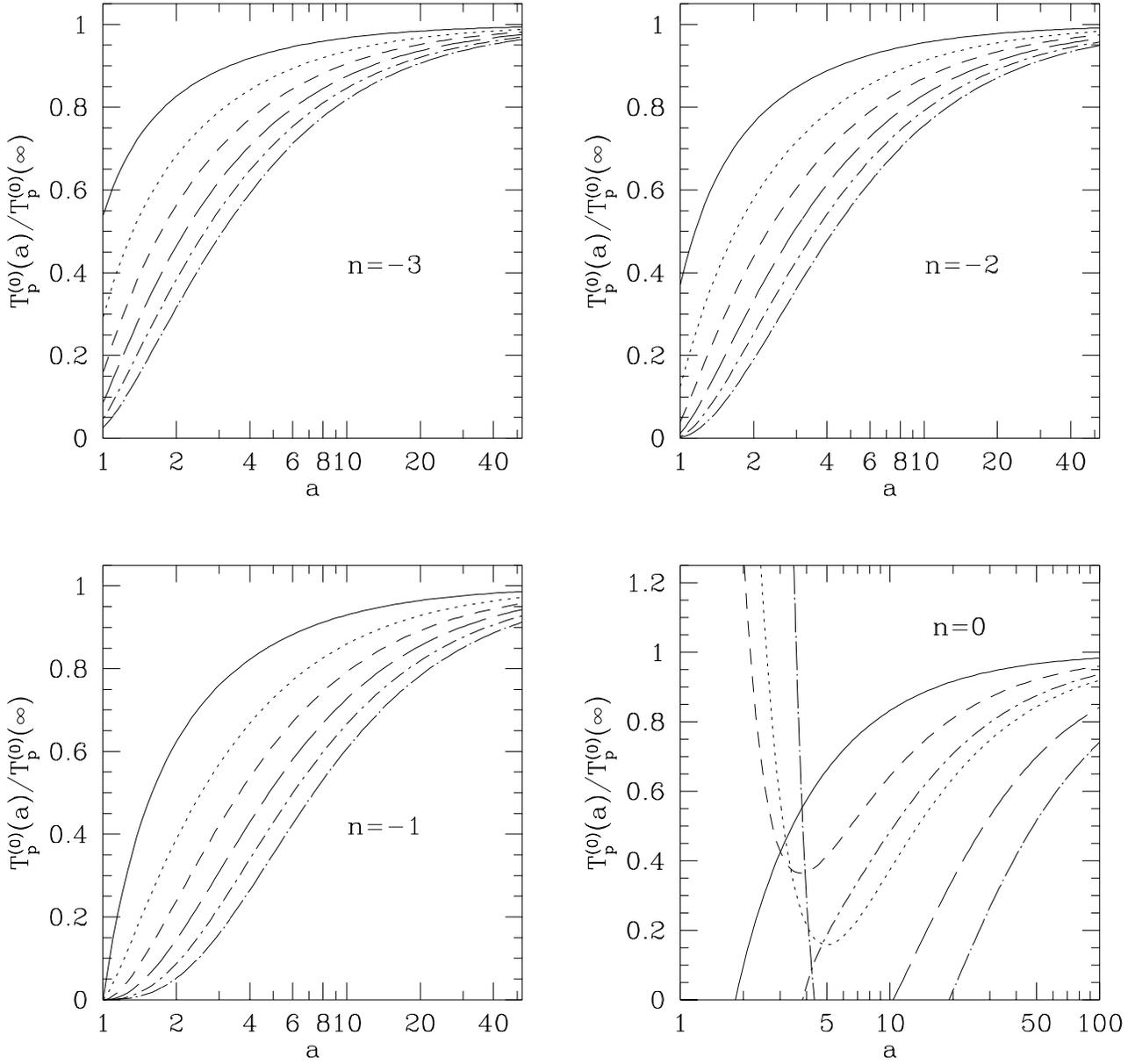}}
\caption{Same as Figure 1, but for the $T_p$ parameters.}
\label{fig2}
\end{figure}

\begin{figure}[t!]
\centering
\centerline{\epsfxsize=18truecm\epsfysize=18truecm\epsfbox{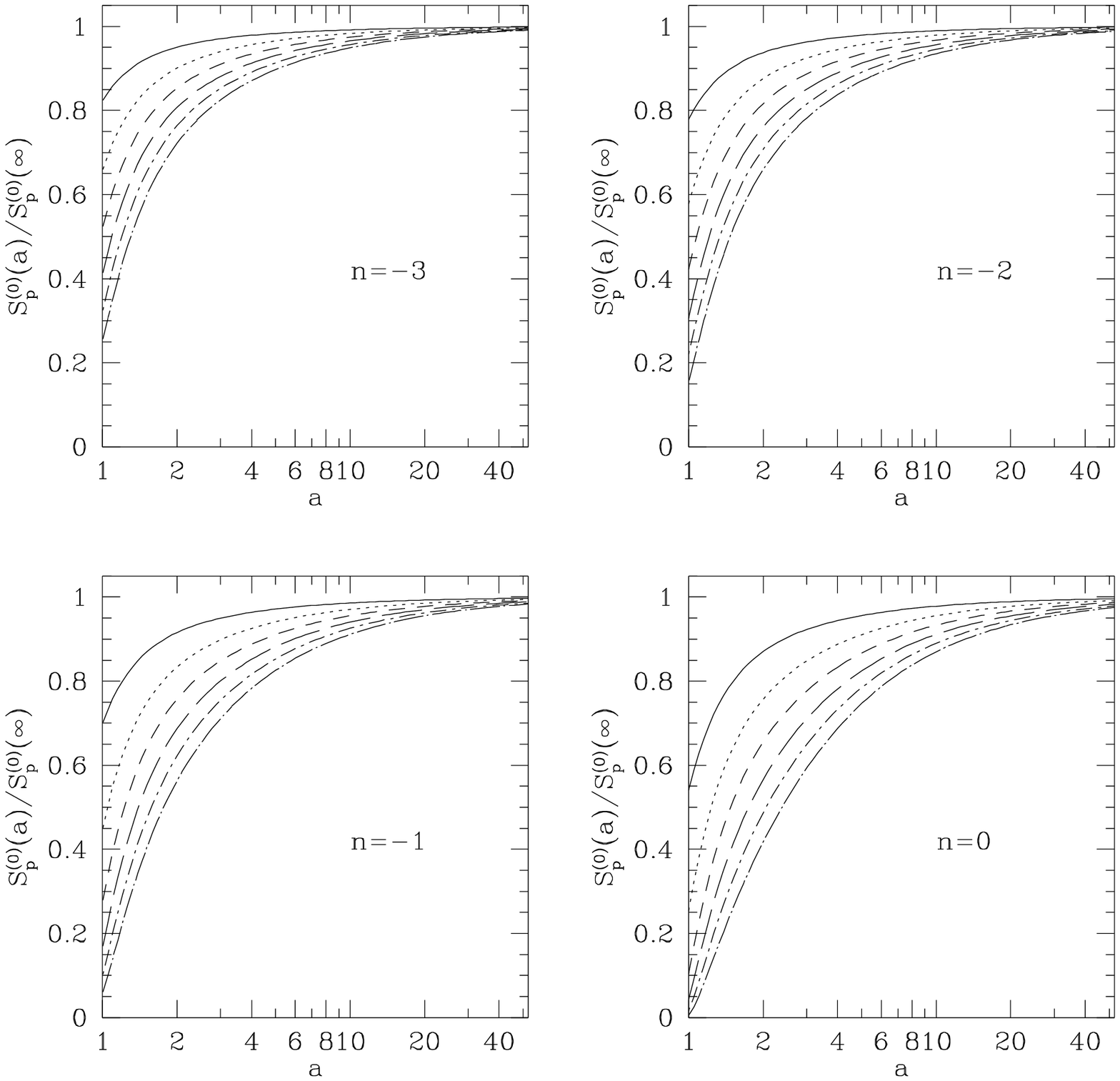}}
\caption{Same as Figure 1, but for initial velocities set as in EDFW,
see Eq.~(\protect{\ref{vel-corr2}}).}
\label{fig3}
\end{figure}

\begin{figure}[t!]
\centering
\centerline{\epsfxsize=18truecm\epsfysize=18truecm\epsfbox{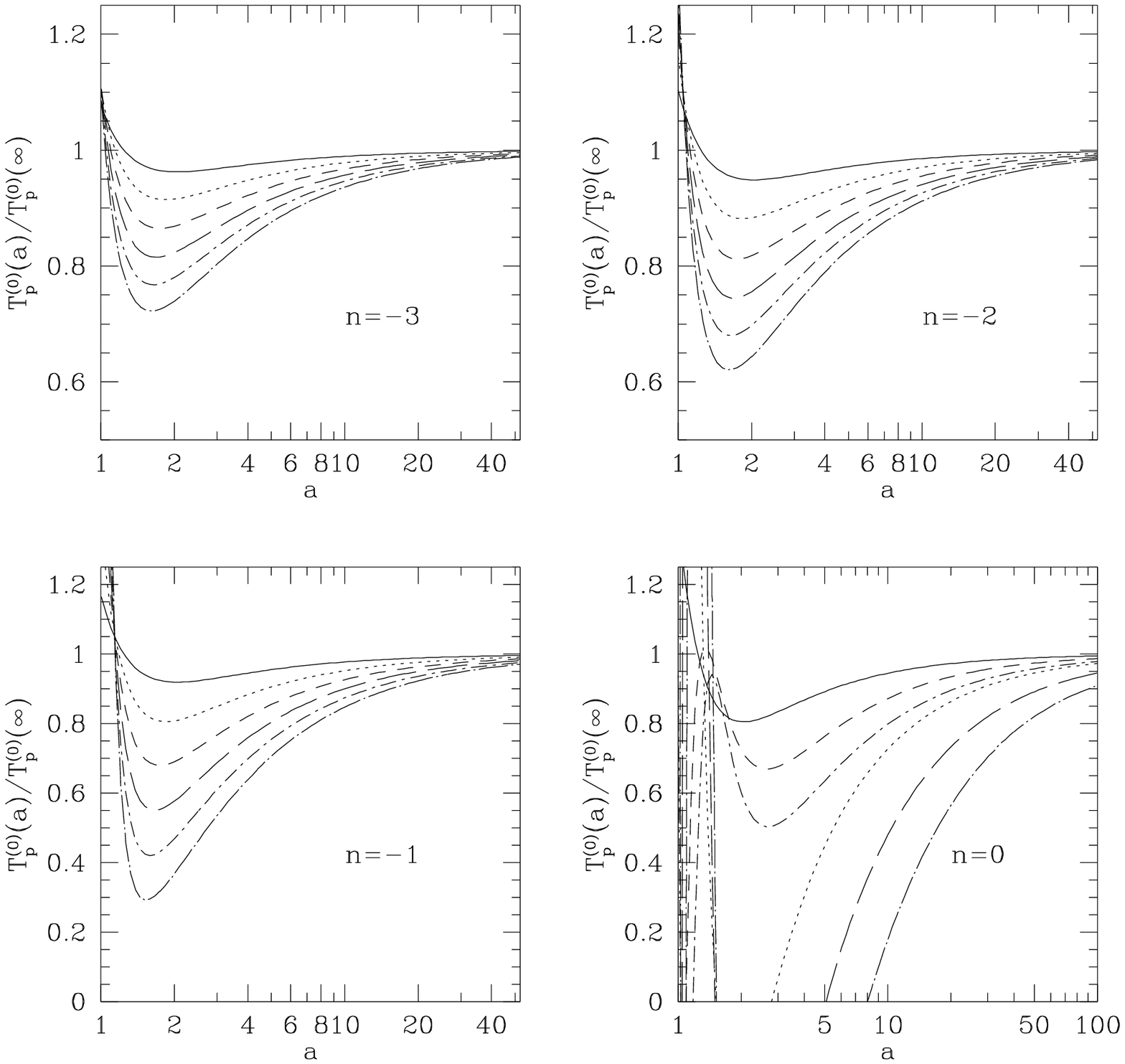}}
\caption{Same as Figure 2, but for initial velocities set as in EDFW,
see Eq.~(\protect{\ref{vel-corr2}}).} 
\label{fig4}
\end{figure}

\begin{figure}[t!]
\centering
\centerline{\epsfxsize=18truecm\epsfysize=18truecm\epsfbox{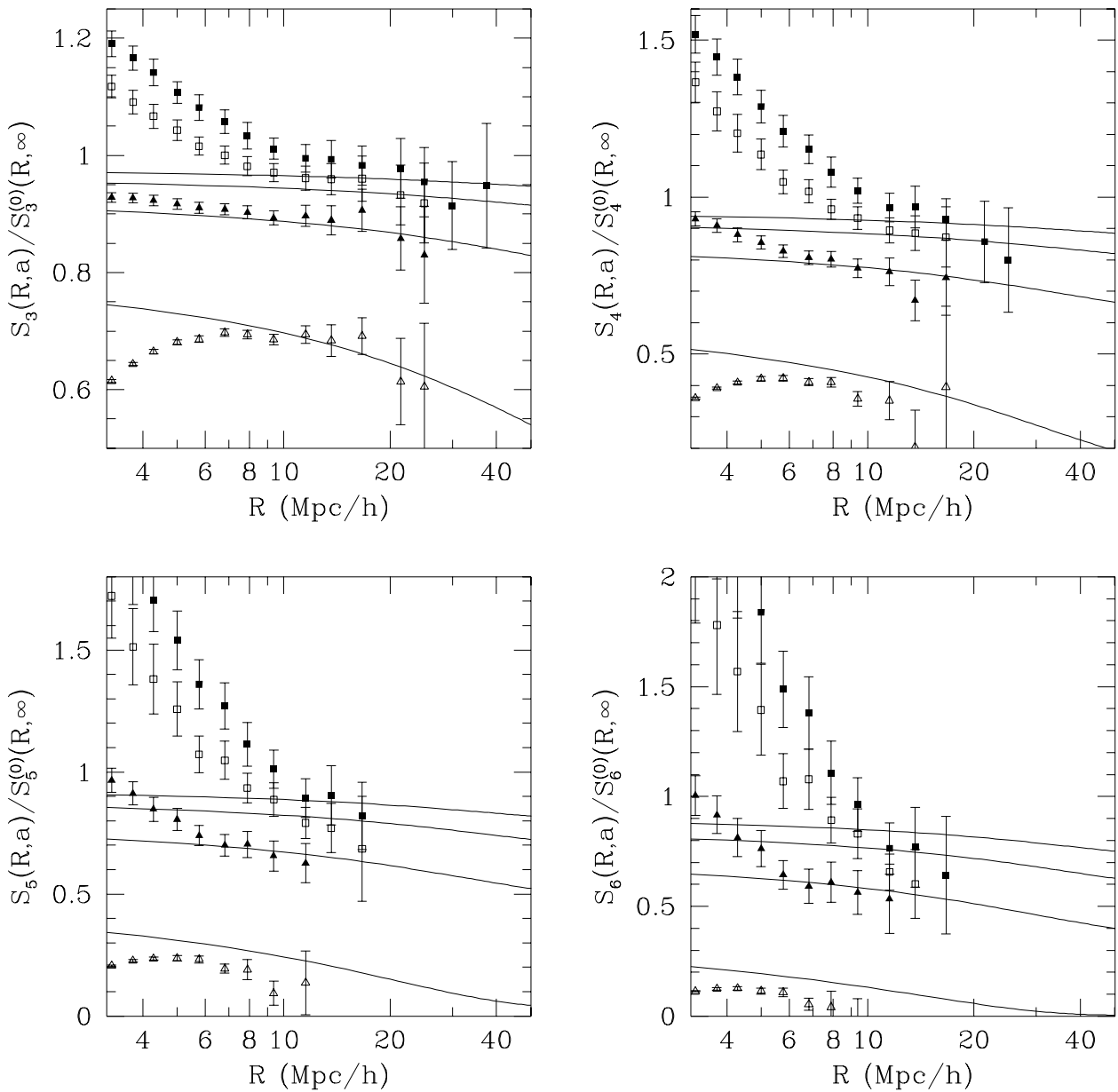}}
\caption{Symbols show the ratio of the $S_p$ parameters at scale factor
$a$ measured in SCDM numerical simulations (Baugh, Gazta\~naga \&
Efstathiou 1995) to their asymptotic
tree-level exact dynamics value as a function of smoothing scale
$R$. Symbols represent $a=1$ (open triangles), $a=1.66$ (filled
triangles), $a=2.75$ (open squares) and $a=4.2$ (filled squares). Error
bars denote the variance of measurements in 10 realizations. Solid
lines correspond to the predictions of transients in tree-level PT,
expected to be valid at large scales.}
\label{fig5}
\end{figure}

\begin{figure}[t!]
\centering
\centerline{\epsfxsize=18truecm\epsfysize=18truecm\epsfbox{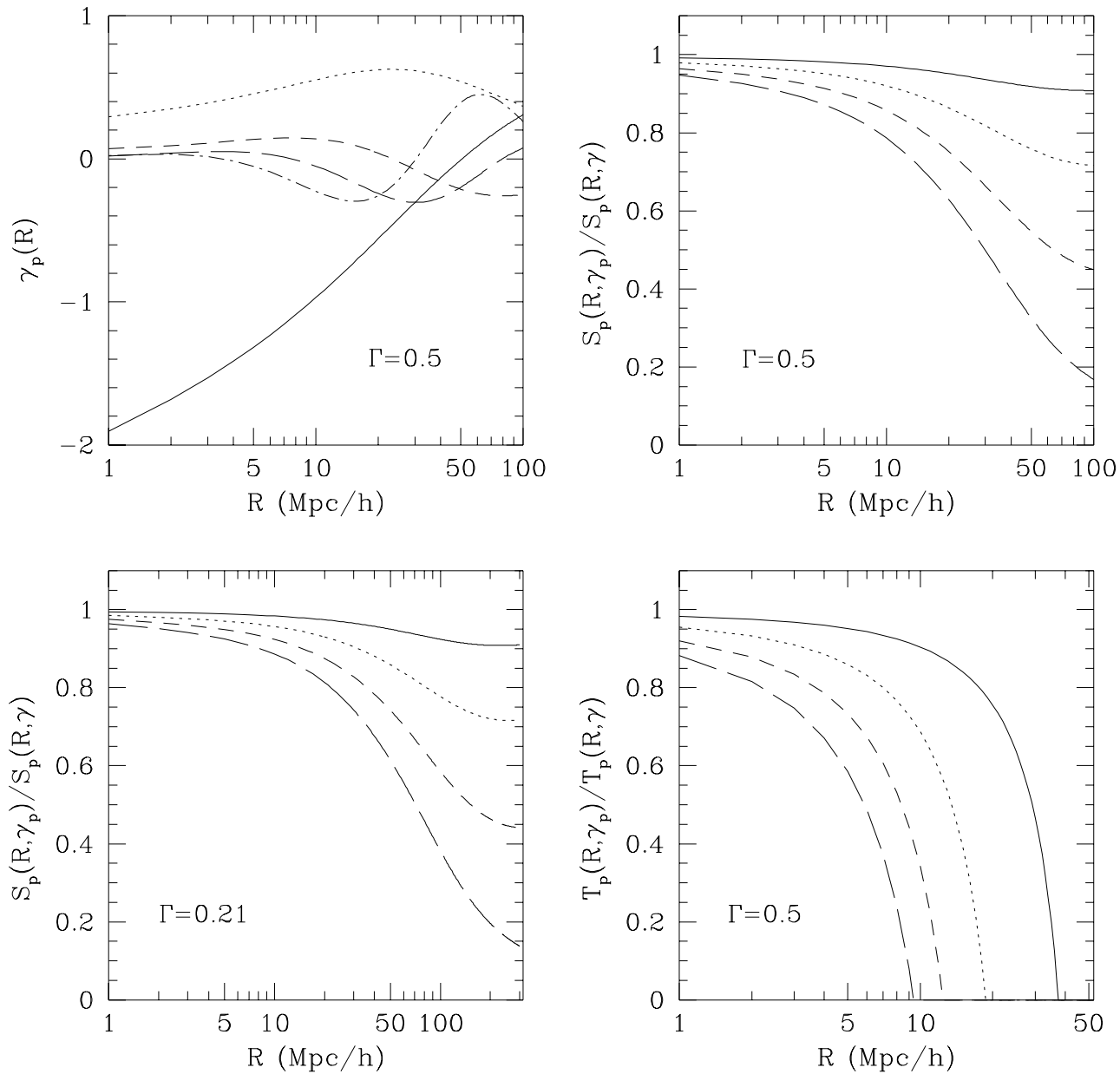}}
\caption{The top left panel shows the $\gamma_p$ parameters as a
function of smoothing scale $R$ for a $\Gamma=0.5$ CDM model. The
solid line corresponds to $n_{\rm eff}(R) \equiv \gamma(R)-3$, and
$\gamma_2$ (dotted line), $\gamma_3$ (short-dashed), $\gamma_4$
(long-dashed), $\gamma_5$ (dot-dashed). The top right panel shows the
ratio of the $S_p$ parameters to those calculated by setting $\gamma_p
=0$ for $p \geq 2$ as a function of scale. Line-styles denote $S_4$
(solid), $S_5$ (dotted), $S_6$ (short-dashed) and $S_7$ (long-dashed).
The left bottom panel shows the
analogous calculation for a $\Gamma=0.21$ CDM model. Right bottom
panel presents a similar result for the $T_p$ parameters in the
$\Gamma=0.5$ CDM model.}
\label{fig6}
\end{figure}

\begin{figure}[t!]
\centering
\centerline{\epsfxsize=18truecm\epsfysize=18truecm\epsfbox{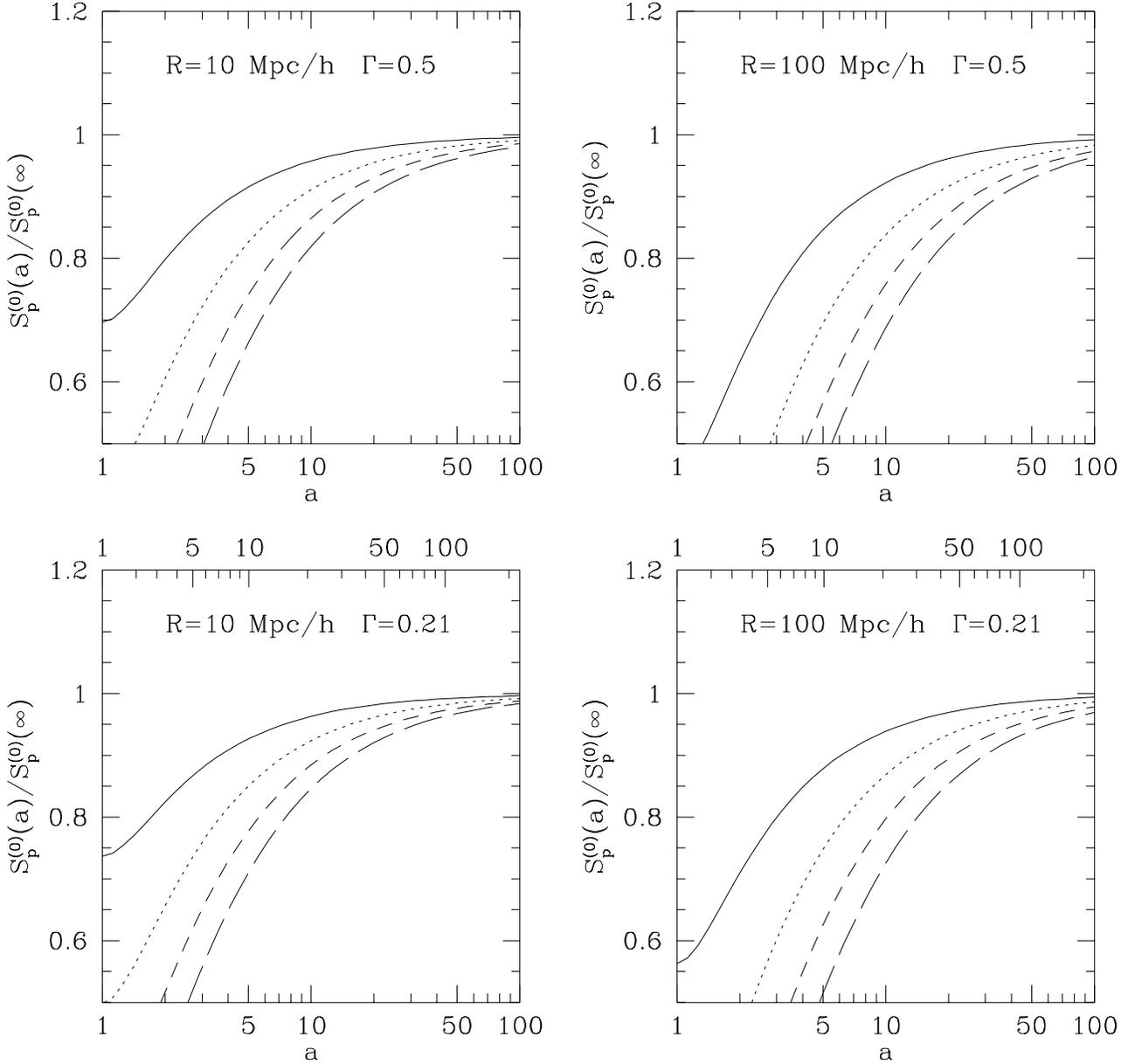}}
\caption{The ratio of the tree-level $S_p$ parameters at scale factor
$a$ to their  asymptotic exact dynamics values for CDM models as a
function of scale factor $a$. Top panels correspond to a $\Gamma=0.5$
CDM model at smoothing scales $R=10,100$ h$^{-1}$Mpc, and bottom
panels show the corresponding results for a  $\Gamma=0.21$
CDM model. Line styles correspond to $S_3$ (solid), $S_4$ (dotted),
$S_5$ (short-dashed), and $S_6$ (long-dashed). The
upper axes in the bottom panels denote the scale factor $a$ for
$\Omega(a)=0.3$, with $\Lambda=0$.}
\label{fig7}
\end{figure}

\begin{figure}[t!]
\centering
\centerline{\epsfxsize=18truecm\epsfysize=18truecm\epsfbox{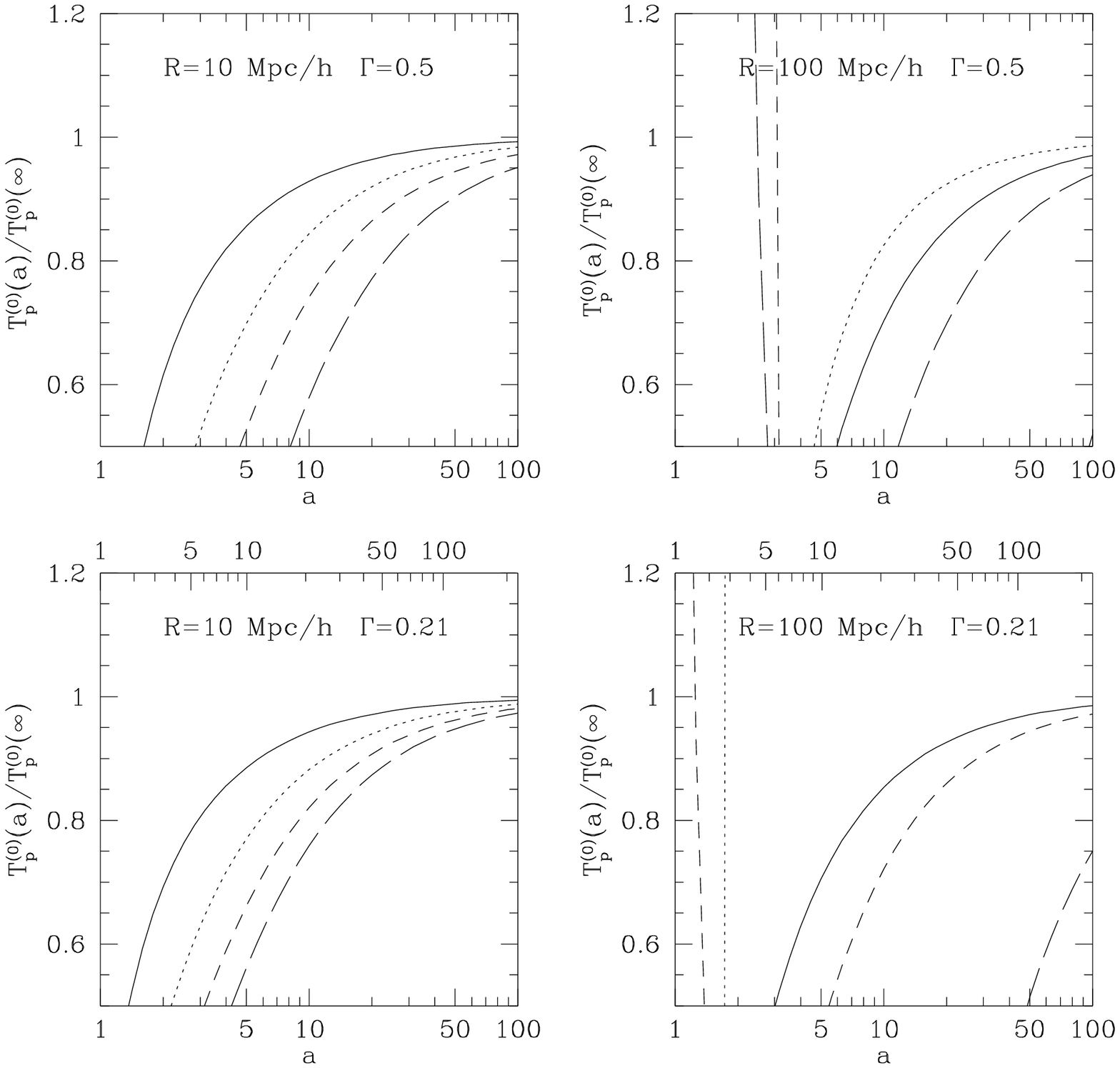}}
\caption{Same as Fig.~\protect{\ref{fig7}}, but for the $T_p$ parameters.} 
\label{fig8}
\end{figure}

\begin{figure}[t!]
\centering
\centerline{\epsfxsize=18truecm\epsfysize=18truecm\epsfbox{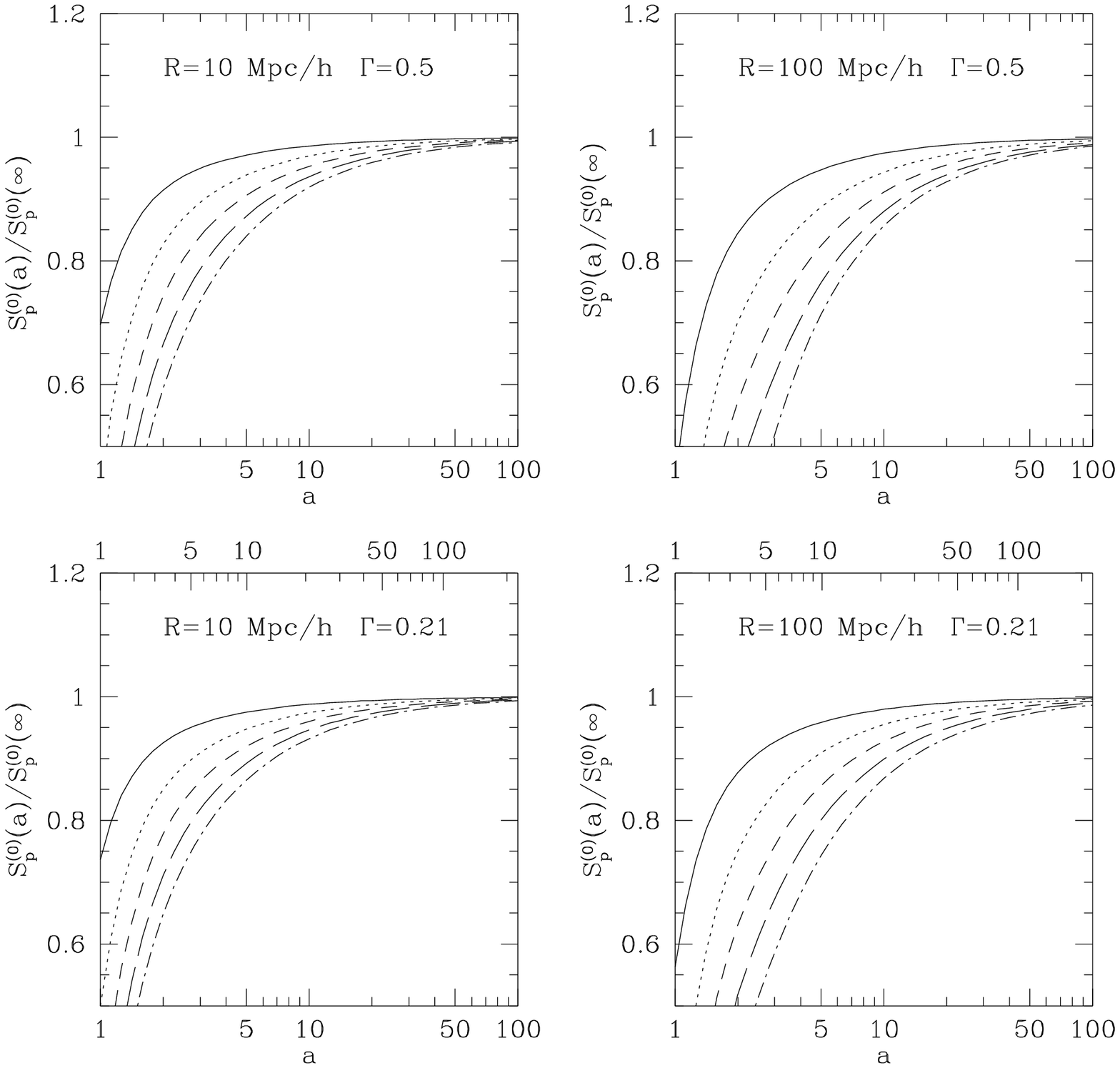}}
\caption{Same as Fig.~\protect{\ref{fig7}}, but for initial velocities
set as in EDFW, see Eq.~(\protect{\ref{vel-corr2}}). The additional
dot-dashed line denotes $S_7$.}  
\label{fig9}
\end{figure}

\begin{figure}[t!]
\centering
\centerline{\epsfxsize=18truecm\epsfysize=18truecm\epsfbox{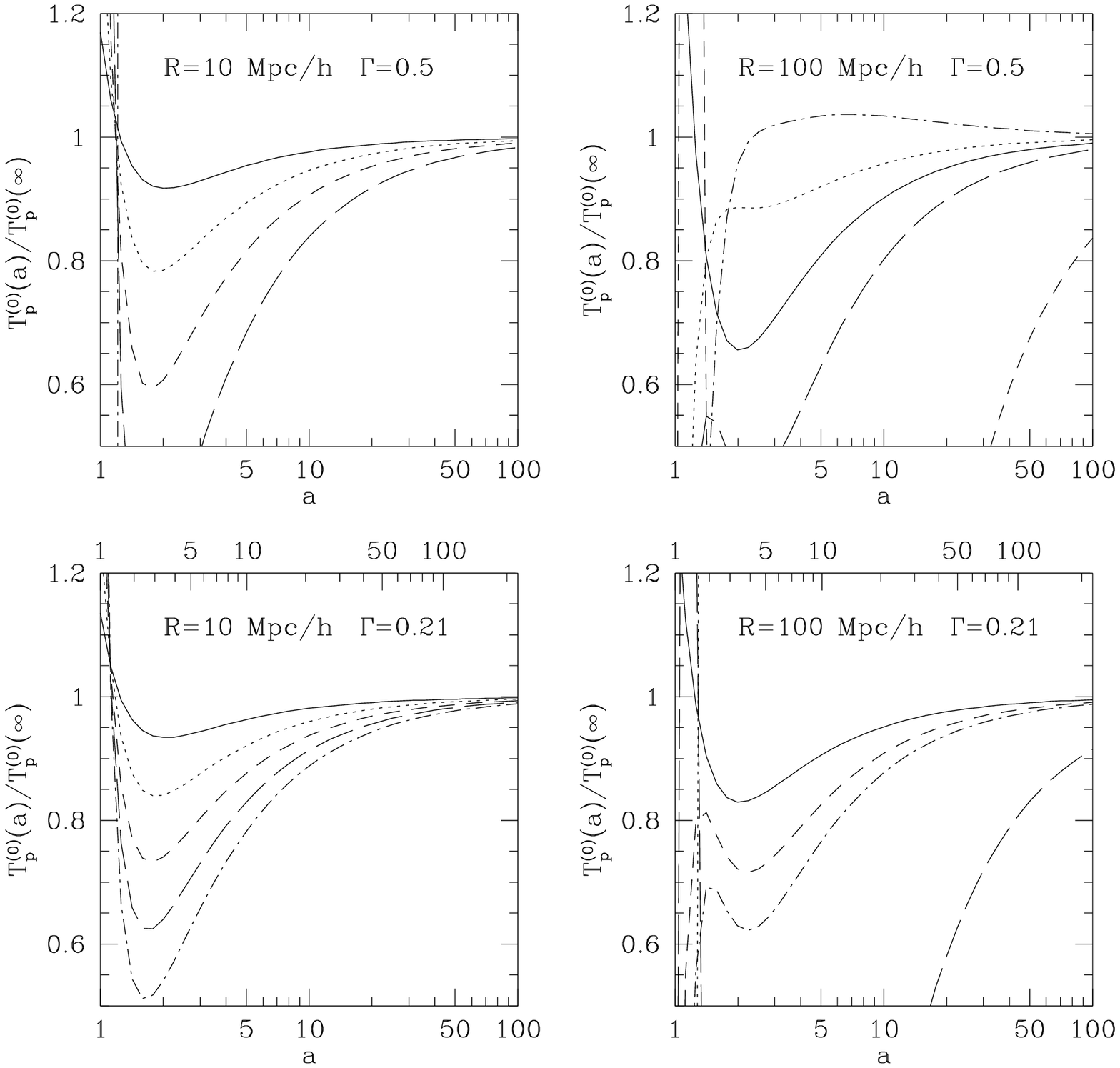}}
\caption{Same as \protect{\ref{fig8}}, but for initial velocities set
as in EDFW, see 
Eq.~(\protect{\ref{vel-corr2}}). The additional dot-dashed line
denotes $T_7$.}  
\label{fig10}
\end{figure}

\begin{figure}[t!]
\centering
\centerline{\epsfxsize=18truecm\epsfysize=18truecm\epsfbox{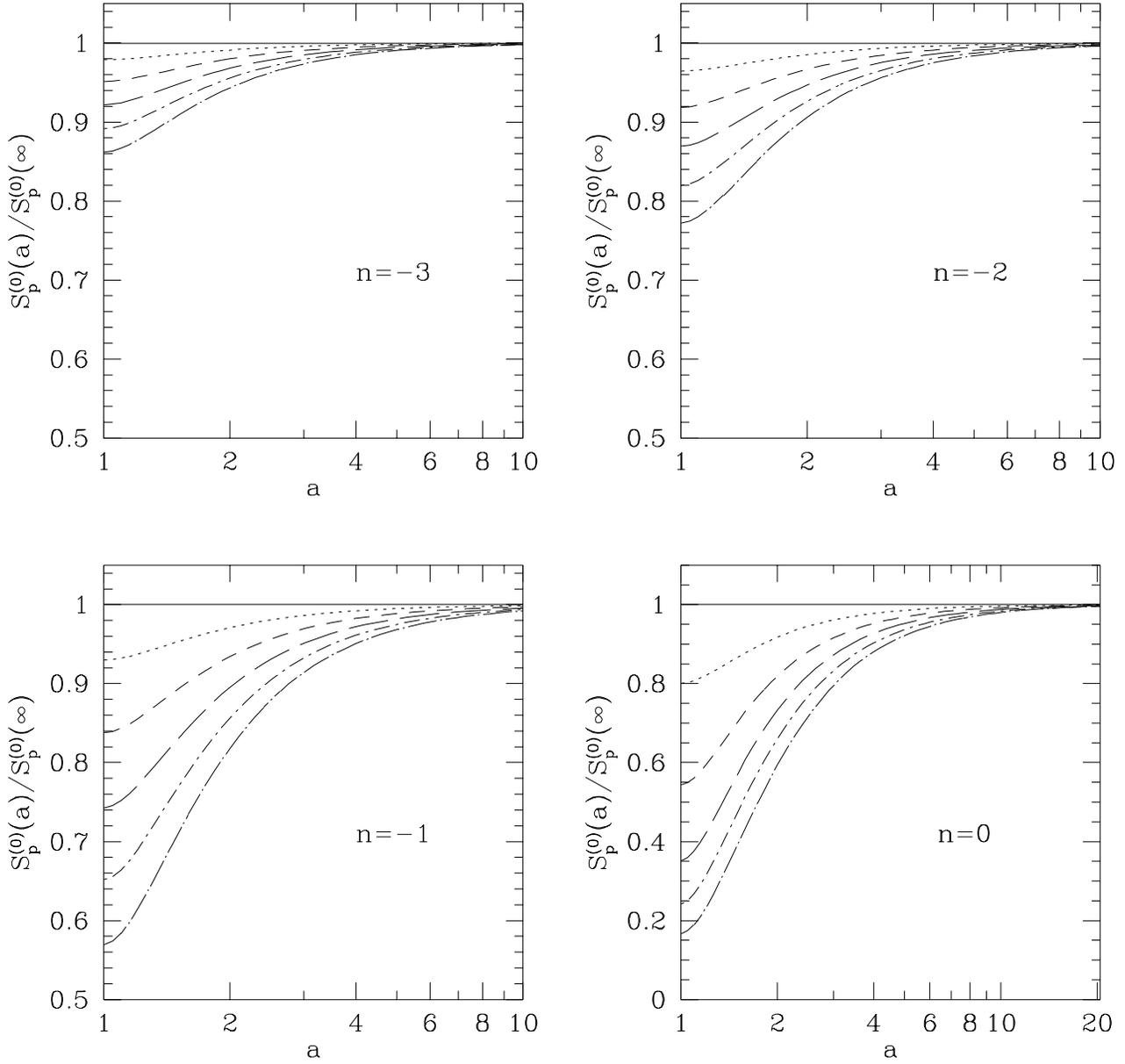}}
\caption{Same as Fig.~\protect{\ref{fig1}}, but for initial conditions set using
second-order Lagrangian perturbation theory. Compare with
Figs.~\protect{\ref{fig1}} and~\protect{\ref{fig3}} (note the
difference in plot scales).} 
\label{fig11}
\end{figure}

\begin{figure}[t!]
\centering
\centerline{\epsfxsize=18truecm\epsfysize=18truecm\epsfbox{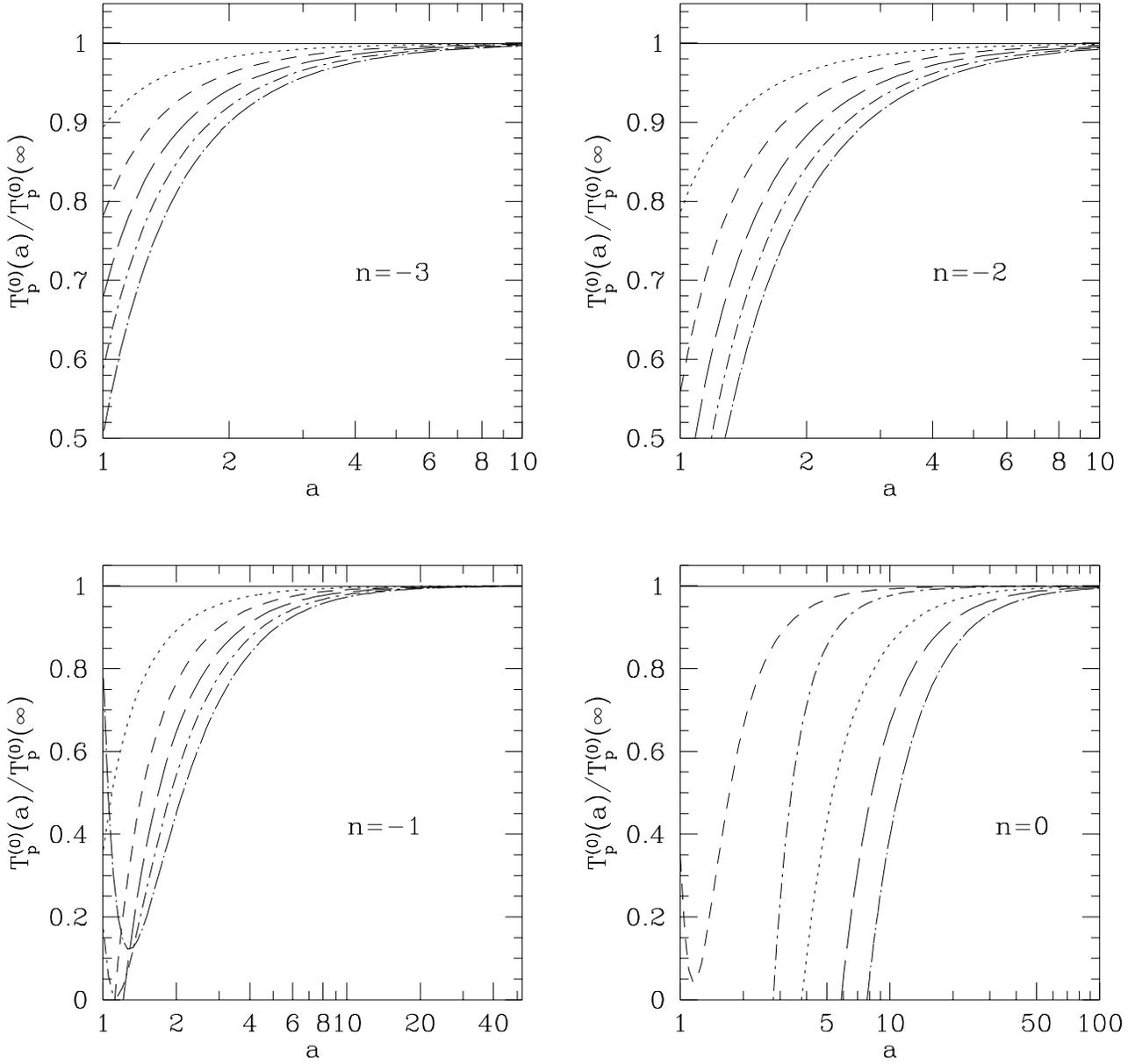}}
\caption{Same as Fig.~\protect{\ref{fig2}}, but for the $T_p$
parameters. Compare with
Figs.~\protect{\ref{fig2}} and~\protect{\ref{fig4}}} 
\label{fig12}
\end{figure}

\end{document}